\documentclass[12pt,a4paper]{article}
\usepackage{latexsym,amsmath,euscript}
\usepackage{slashed, pstricks}
\renewcommand{\a}{{\alpha}}

\renewcommand{\b}{{\beta}}
\newcommand{\ga}{{\gamma}}
\newcommand{\dl}{{\delta}}
\newcommand{\eps}{{\epsilon}}

\newcommand{\ka}{{\kappa}}
\newcommand{\la}{{\lambda}}
\newcommand{\m}{{\mu}}

\newcommand{\sig}{{\sigma}}

\newcommand{\om}{{\omega}}

\newcommand{\pa}{{\partial}}

\newcommand{\La}{{\Lambda}}

\newcommand{\ds}{\displaystyle}
\newcommand{\ssc}{\scriptstyle}
\newcommand{\dr}{\textnormal{\bf d}}

\makeatletter 
\renewcommand{\section}{\@startsection{section}{1}{\z@}{-6ex plus -1ex minus
    -.2ex}{2.3ex plus .2ex}{\large\bf}}
\renewcommand{\subsection}{\@startsection{subsection}{2}{\z@}{-3.25ex plus -1ex
    minus -.2ex}{1.5ex plus .2ex}{\normalsize\bf}}
\renewcommand{\@dotsep}{200} 
\makeatother

\numberwithin{equation}{section}
\renewcommand{\theequation}{\arabic{section}.\arabic{equation}}

\begin{document} 
\thispagestyle{empty}

{\bf\Large %
  \centerline{Quantization of the open string on plane-wave limits of} %
  \centerline{$\boldsymbol{dS_n \times S^n}$ and non-commutativity outside
    branes}%
}

\vskip 40pt
\begin{center}
{\large G. Horcajada and F. Ruiz Ruiz}\\[3pt]
{\it Departamento de F\'{\i}sica Te\'orica I, Universidad
     Complutense de Madrid,\\
     28040 Madrid, Spain}
\vskip 40pt
\date{}
\end{center}

\begin{abstract}
  \noindent
  The open string on the plane-wave limit of $dS_n\times S^n\,$ with constant
  $B_2$ and dilaton background fields is canonically quantized. This entails
  solving the classical equations of motion for the string, computing the
  symplectic form, and defining from its inverse the canonical commutation
  relations.  Canonical quantization is proved to be perfectly suited for this
  task, since the symplectic form is unambiguously defined and non-singular.
  The string position and the string momentum operators are shown to satisfy
  equal-time canonical commutation relations. Noticeably the string position
  operators define non-commutative spaces for all values of the string
  world-sheet parameter $\sig$, thus extending non-commutativity outside the
  branes on which the string endpoints may be assumed to move.  The Minkowski
  spacetime limit is smooth and reproduces the results in the literature, in
  particular non-commutativity gets confined to the endpoints.

\end{abstract}

\section{Introduction}

Solutions to the Einstein equations in general relativity have been known for
a long time to have plane waves as limits~\cite{Penrose}. These limits, known
as Penrose limits, give a plane wave spacetime approximation for the full
spacetime along a null geodesic. This observation led in the sixties and
seventies to a detailed study of the geometric properties of plane-wave
metrics and of matter fields defined on them~\cite{Stephani}. Already within
string theory, it soon became clear that higher-dimensional plane waves give
exact solutions to string theory, provided the Kalb-Ramond and dilaton fields
satisfy certain conditions~\cite{Amati}~\cite{Horowitz-Steif-1}.  The
generalization of the Penrose limiting procedure relating higher dimensional
plane waves with more complicated solutions to string theory~\cite{Gueven}
further triggered the interest in such space-times.

By now, there is a very extensive literature on plane waves in string theory.
Motivated by the fact that \hbox{$\,AdS_4\times S^7\,$} and
\hbox{$\,AdS_7\times S^4\,$} are solutions to M-theory and \hbox{$AdS_5\times
  S^5$} is a solution of IIB supergravity, and by the AdS/CFT correspondence,
special attention has has been given the Penrose limit~\cite{KG,Blau,Blau-2}
\begin{equation}
    AdS_k\!\times\! S^n~\,pp\textnormal{-limit:}\quad 
    ds^2= - \,dx^+ dx^- - m^2\,\mathbf{x}^2_{k+n-2}\>{(dx^+)}^2\! 
      + d\mathbf{x}^2_{k+n-2} 
\label{pp-AdSxS}
\end{equation}
of $AdS_k\times S^n$ spaces. Two milestones in this regard are (i) the
quantization~\cite{Metsaev} of the R-R sector of the closed superstring on
this background for $k=n=5$, and (ii) the derivation of its spectrum from that
of $U(N)$ ${\cal N}=4$ super Yang-Mills theory~\cite{Berenstein}. The interest
has extended also to type IIB superstring models in 6 dimensions~\cite{Russo}
describing generalizations of the Nappi-Witten model. As a matter of fact, the
Nappi-Witten model~\cite{Nappi-Witten} is itself the Penrose limit of
$AdS_2\times S^2$. 
There has been as well interest on strings on 4-dimensional homogenous
plane-wave backgrounds~\cite{Papadopoulos}. These have the
form~(\ref{pp-AdSxS}) with $m^2$ replaced by a function $\,C|x^+|^{-2}$ and,
for different values of the constant $C$, occur as the Penrose limit of the
spatially flat FRW metric, near horizon regions of \emph{Dp}-brane backgrounds
and fundamental strings backgrounds~\cite{Blau-2,Fuji}

In this paper we consider quantization of the open string on the Penrose limit
of $dS_n\times S^n$ with non-zero constant 2-form $B_2$.  To date, no
background \emph{p}-forms have been found that support $dS_n\times S^n$ as a
solution to IIB supergravity. Yet there are indications that de Sitter space
may occur in type IIA theories~\cite{Silverstein}. In any case, the Penrose
limit of $dS_n\times S^n$ is an exact solution of string theory in the
critical dimension~\cite{Horowitz-Steif-1}. There are other motivations for
taking de Sitter space-time: its ``apparent'' simplicity when it comes to
quantum gravity~\cite{Witten-deSitter}, the dS/CFT
correspondence~\cite{Strominger} and the fact that the non-existence of a
positive conserved energy indicates that there cannot be unbroken
supersymmetry, so it seems a good starting point to go down in the number of
supersymmetries.  The motivation for taking $B_2\neq 0$ comes from an interest
in understanding non-commutativity in relation with gravity. As is well-known,
string theory gives explicit realizations of non-commutative spaces. The
simplest example is provided by an open string in Minkowski spacetime with
endpoints moving on a \emph{D}-brane on which a magnetic field is defined:
upon quantization, the string position operators generate a non-commutative
space along the brane~\cite{SW,Chu-Ho-1,SJ}. Since non-commutativity is
postulated as a candidate to reconcile quantum mechanics with general
relativity~\cite{DFR}, and the low energy limit of string theory includes
general relativity, it seems natural to explore the non-commutativity/gravity
connection within string theory.  One way to push forward this approach is to
examine non-commutativity for plane wave backgrounds.  As a matter of fact,
this program has already started for the open string on plane-wave limits of
$AdS_n\times S^n$. In 10 dimensions with a constant non-zero $B_2$ in
ref.~\cite{Chu-Ho-3}, and in 4 dimensions with a Nappi-Witten 2-form
in~\cite{Dolan-Nappi}. In both instances, the string endpoints define
non-commutative spaces. Here we investigate non-commutativity for the Penrose
limit of $dS_n\times S^n$.

More precisely, we will quantize the open string interacting through a
plane-wave metric
\begin{equation}
  ds^2 = -\,dx^+\!\, dx^- + m^2\,\big[{(x^1)}^2 -{(x^2)}^2\,\big]\,{(dx^+)}^2 
  + \sum_{i=1}^2{(dx^i)}^2 + \sum_{a=3}^{D-2}{(dx^a)}^2
\label{pp-metric}
\end{equation}
and constant antisymmetric and dilaton fields
\begin{equation}
  B_{ij} =\eps_{ij}B \qquad  B_{ab}= 0 \qquad \Phi=\Phi_0\,.
\label{dilaton}
\end{equation}
It will come out that the string position operators $\,X^1(\tau,\sig)\,$ and
$\,X^2(\tau,\sig^\prime)\,$ do not commute for arbitrary $\sig$ and
$\sig^\prime$. This is in contrast with the results available so far for open
strings on $AdS_n\times S^n$ plane-wave limits supported by a non-zero
$B_2$~\cite{Chu-Ho-1,Chu-Ho-3,Dolan-Nappi}, for which non-commutativity is
restricted to the brane manifold on which the string endpoints move. Our
results are consistent with those in the literature for Minkowski
spacetime~\cite{Chu-Ho-1}, since the latter are recovered in the limit $m\to
0$, and in particular non-commutativity gets confined to the string endpoints.

We will work in light-cone and conformal gauges. The paper is organized as
follows. In Section 2 we derive the equations of motion for the classical
string and solve them. The solution turns out to be an infinite sum over
modes, with a highly non-trivial dependence on the parameter $m$. As compared
to the open string in Minkowski spacetime, two important differences are
encountered. The first one is that the string has a finite number of
non-oscillating degrees of freedom associated to modes exponentially growing
and decaying in $\tau$. The second one is that the string total momentum is
not an independent degree of freedom but receives contributions from all the
modes.  In Section 3 the string is canonically quantized. This is done by
calculating the symplectic form and then using it to find the commutation
relations for the operators associated to all the string modes. The symplectic
form is unambiguous and non-singular, not being necessary to provide
additional constraints or to modify its definition so as to fix the
commutators. As a check it is shown that the string momentum and the string
position operators satisfy equal-time canonical commutation relations. Section
4 shows that the string position operators $X^1$ and $X^2$ do not commute for
arbitrary values of $\sig$ and $\sig^\prime$, thus defining non-commutative
waves fronts. In Section 5 we find the eigenstates and spectrum of the
hamiltonian. Section 6 contains our conclusions. We have included two
Appendices with some of the details of the calculations of Sections 2 and 4.

\section{The classical string}

Due to its length, this section is divided into five parts. In the first one,
we study the background metric~(\ref{pp-metric}). The second subsection
contains the derivation of the equations of motion and of the boundary
conditions for the classical open string in the
background~(\ref{pp-metric})-(\ref{dilaton}).  The equations of motion are
solved in the thir part, where expressions for the string coordinates as sums
over modes ready to be quantized are found. The fourth subsection presents a
brief discussion of the string center of mass coordinates and the string total
momentum. Finally, in the fifth part we discuss the case $\,m^2\ka^2\!\ll\!1$.

\subsection{The background as the Penrose limit of $\boldsymbol{dS_n\times
    S^n}$}

The metric~(\ref{pp-metric}) is the Penrose limit of $\,{dS}_2\times
{S^2}\times \textnormal{E}^{D-4}$, with ${\rm E}^{D-4}$ euclidean space in
$D-4$ dimensions.  Although well-known, let us very briefly check this point.
Consider $k$-dimensional de Sitter space-time times an $n$-sphere,
$\,dS_k\times S^n$, both or radius $\ell$. Its metric can be written as
\begin{equation}
     ds^2 = \ell^2\, \Big[ - (1-\rho^2)\,dt^2
               + \frac{d\rho^2}{1-\rho^2} + \rho^2\,d\Omega^2_{k-2}
               + (1-r^2)\,d\chi^2 + \frac{dr^2}{1-r^2} 
               + r^2\,d{\Omega^\prime}^2_{n-2}\,\Big]\,,
\label{dSxS}
\end{equation}
where $d\Omega^2_{k-2}$ and $\,d{\Omega^\prime}^2_{n-2}$ are the round metrics
on the unit $(k-2)$ and $(n-2)$-spheres. Consider now, as in the anti-de
Sitter case~\cite{Berenstein}, the trajectory along $\chi$ in the vicinity of
$\rho=r=0$. Making the changes $\,u^\pm=t\pm\chi$, rescaling
\begin{equation}
    u^+=x^+ \quad u^-=\frac{x^-}{\ell^2} \quad 
    \rho=\frac{\bar{\rho}}{\ell} \quad r=\frac{\bar{r}}{\ell} 
    \quad \textnormal{with~~}\ell\to\infty\,,
\label{pp-limit}
\end{equation}
and introducing a mass scale $x^+\to 2mx^+,~x^-\to x^-/2m$, one arrives at
\begin{equation}
     dS_k\!\times\! S^n~\,pp\textnormal{-limit:}\quad
    ds^2_{pp} = - \,dx^+ dx^-\! + m^2 \big( \mathbf{x}^2_{k-1} 
      - \mathbf{y}^2_{n-1} \big) \,{(dx^+)}^2 
      + d\mathbf{x}^2_{k-1}\! + d\mathbf{y}^2_{n-1}.
\label{pp+-}
\end{equation}
Here cartesian coordinates
\hbox{$\mathbf{x}_{k-1}\!=(\bar{\rho},\Omega_{k-2})\,$} and
$\,\mathbf{y}_{n-1}\!=(\bar{r},\Omega_{n-2})$ have been introduced.
Backgrounds
\begin{equation*}
  ds^2 = ds^2_{pp}+ds^2({\rm E}^{D-n-k}) \qquad
  H_3=dB_2=A_{ij}(x^+)dx^i\wedge dy^j
\end{equation*}
are solutions to all orders in $\alpha^\prime$ for the bosonic/fermionic
string in $D=26/10$ provided $A_{ij}$ satisfies the
condition~\cite{Horowitz-Steif-1}
\begin{equation*}
   4\,m^2(n-k)=A_{ij}\,A^{ij} \,.
\end{equation*}
$H_3$ vanishes for $k=n$, in which case one may take
$\,B_2\!=B_{ij}\,dx^i\wedge dy^j$, with $B_{ij}$ constant. The
metric~(\ref{pp-metric}) is recovered for $k=n=2$ and is non-singular, meaning
it is geodesically complete. The results in this paper are trivially extended
to the case $k=n=5$.

It is important to note the positive sign in front of $\mathbf{x}^2_{k-1}$ in
the metric coefficient $g_{++}$ in eq.~(\ref{pp+-}). This has its origin in
the fact that we have started with de Sitter space-time, rather than anti-de
Sitter, and implies that the metric~(\ref{pp+-}) does not admit a conserved
positive energy. To understand this we recall that in de Sitter space there is
no positive conserved energy since there is no generator of its isometry
group, $SO(1,d)$, which is timelike everywhere. In the
coordinates~(\ref{dSxS}), the generator $\pa/\pa t$ is timelike for $\rho<1$,
but vanishes at the event horizon $\rho=1$.  Hence, $\pa/\pa t$ and its
associated hamiltonian can only be used to define time evolution in the region
$\,0\leq \rho \leq 1$ within the event horizon. Upon forming $dS_n\times S^n$
and taking the Penrose limit, this implies that for the metric~(\ref{pp+-})
the sign of the energy depends on the sign of $\mathbf{x}^2_{k-1} -
\mathbf{y}^2_{n-1}$. This is a property of the background considered.

\subsection{Classical action, field equations and momenta}

Our starting point is the bosonic part of the classical action 
\begin{equation*}
  S=\frac{1}{4\pi\alpha'}\int\! d\tau\,d\sigma\, \Big(
       \sqrt{-\gamma}\,\gamma^{rs} \, G_{\mu\nu} \,
       \partial_r X^{\mu}\,\partial_s X^\nu
     + \eps^{rs} B_{\mu\nu}\, \partial_r X^{\mu}\,\partial_s X^\nu
     + \alpha'\sqrt{-\gamma}\, R \,\Phi \Big)
\end{equation*}
for the open string on the $D$-dimensional background
$G_{\m\nu}(X),\,B_{\mu\nu}(X),\,\Phi(X)$ in
eqs.~(\ref{pp-metric})-(\ref{dilaton}). Greek letters $\m,\nu,\ldots$ denote
spacetime indices, while lower case letters $r,s,\ldots$ from the end of the
Roman alphabet denote world-sheet indices. Here $\ga_{rs}$ is the metric on
the string world-sheet, $R$ its scalar curvature and $\eps^{rs}$ is defined by
$\,\eps^{01}\!=1$. As usual the world-sheet coordinates $\tau$ and $\sigma$
take values on the intervals $\,\,-\infty<\tau<\infty\,\,$ and $\,\,0\leq
\sigma\leq\pi$. We are using units in which string coordinates have dimensions
of length and $\tau,\sig$ are dimensionless. From now on we will use capital
case letters $X'\textnormal{s}$ for the string coordinates.

If wished, the string endpoints may be assumed to lie on a \emph{Dp}-brane on
which a magnetic field $F_{ij}$ lives\footnote{$p$ is 1 for
  $\,k\!=\!n\!=\!2\,$ in~(\ref{pp+-}) and 4 for $k\!=\!n\!=\!5$.}.  This
amounts to adding to the action a term
\begin{equation*}
    \dl S = \frac{1}{2\pi\alpha^\prime} \int \! d\tau\, 
       A_{i}\, \pa_\tau\! X^{i}\,
         {\Bigg\vert}^{\sigma=\pi}_{\sigma=0} \,,
\end{equation*}
with $\,A_i(X)\,$ the $\,U(1)\,$ gauge field on the brane. If this term is
included in the action, the analysis in this paper goes through with the only
difference that the field $B_{ij}$ must be replaced by the Born-Infeld field
strength $\,{\cal B}_{ij}\!=B_{ij}-F_{ij}$, where $F_{ij}$ is the $\,U(1)\,$
field strength on the brane.

The string action has three world-sheet symmetries. We will fix one of them by
working in light-cone gauge~\cite{Horowitz-Steif-2}
\begin{equation*}
   X^+ = \ka\tau \,,
\label{light-cone}
\end{equation*}
with $\ka$ a parameter with dimensions of length. The other two will be fixed
by choosing conformal gauge
\begin{equation*}
      h^{rs} = \sqrt{-\ga} \,\ga^{rs} = {\rm diag}\,(-1,+1) \,.
\end{equation*}
In this gauge, the classical action becomes 
\begin{equation*}
      S = \int\! d\tau\> L\,,
\end{equation*}
where the lagrangian $L$ is given by
\begin{align*}
  L & = p_-\, \pa_\tau x^- 
     \\[3pt]
  & - \frac{1}{4\pi\a^\prime} \int_0^\pi \! d\sig~ \Big\{ 
        m^2\ka^2 \big[ \,{(X^1)}^2 - {(X^2)}^2\,\big] 
       + \big(\pa_\tau X^i\big)^2 + \big(\pa_\tau X^a\big)^2 
     \\[3pt]
    & \hphantom{\frac{1}{4\pi\a^\prime} ~~~}
    - \big(\pa_\sig X^i\big)^2 
                - \big(\pa_\sig X^a\big)^2\, 
      - 2\,B\,\big[ \pa_\tau X^1\, \pa_\sig X^2 - \pa_\sig X^1\,\pa_\tau X^2
              \big] \Big\} \,, 
\end{align*}
with 
\begin{equation*}
  p_-= - \,\frac{\ka}{4\a^\prime}
\end{equation*}
the momentum conjugate to $x^-(\tau)$, defined~\cite{Polchinski} as the
average over $\sig$ at a given $\tau$ of $X^-(\tau,\sig)$
\begin{equation*}
    x^-(\tau) = \frac{1}{\pi}\int_0^\pi\!d\sig~ X^-(\tau,\sig)\,.
\end{equation*}
Here we have reserved the subscript $i$ for the 1 and 2 directions, while $a$
runs from 3 to $D-2$, a convention that we will follow from now on.  

The field equations and boundary conditions are obtained by varying the action
with respect to $X^i$ and $X^a$. They take the form
\begin{align}
     & \Box X^1 + m^2\ka^2 X^1=0 \label{X1-eq-lc}\\[3pt]
     & \Box X^2 - m^2\ka^2 X^2=0 \label{X2-eq-lc} \\[3pt]
     & \Box X^a = 0\,, \label{Xa-eq-lc}
\end{align}
with $\Box=-\pa_\tau^2+\pa_\sig^2$ the 2-dimensional d'Alambertian,
and 
\begin{eqnarray}
    & {\pa_\sig X^1 - B\,\pa_\tau X^2 \Big\vert}_{\sig=0,\pi} = 0 
    & \label{X1-bc}\\[3pt]
    & {\pa_\sig X^2 + B\,\pa_\tau X^1 \Big\vert}_{\sig=0,\pi}=0 & 
        \label{X2-bc}\\[3pt]
   & {\pa_\sig X^a\Big\vert}_{\sig=0,\pi}=0\,. & \label{Xa-bc}
\end{eqnarray}

To quantize the theory we will need the momenta. In our case, these are given
by 
\begin{align}
  p_- & = - \frac{\ka}{4\a^\prime} \label{Pv} \\
  P_i & = \frac{1}{2\pi\a^\prime}~\big(\pa_\tau X^i 
          - B\,\eps_{ij}\pa_\sig X^j\big) \label{Pi} \\
  P_a &= \frac{1}{2\pi\a^\prime}~\pa_\tau X^a \,. \label{Pa}
\end{align}
In terms of them, the lagrangian $L$ can be written as
\begin{equation*}
   L= -\, p_-\, \pa_\tau x^- + \int_0^\pi \! d\sig\> \big[  
        - \big(  P_i\, \pa_\tau X^i + P_a\, \pa_\tau X^a  \big) 
        + {\cal H}\,\big]\,,
\end{equation*}
where the hamiltonian density ${\cal H}$ has the form
\begin{align}
   4\pi\a^\prime\, {\cal H} 
       & =  \big( 2\pi\a^\prime P_i + B\epsilon_{ij}\, \pa_\sig X^j\big)^2
           + \big(2\pi\a^\prime P_a \big)^2 + {(\pa_\sig\,X^i)}^2
           + {(\pa_\sig\,X^a)}^2 \nonumber \\[3pt]
   & - m^2\ka^2 \,\big[ {(X^1)}^2 -{(X^2)}^2\,\big] \,. 
   \label{ham-den}
\end{align}
We note that ${\cal H}$ is not positive definite because of the negative sign
in front of ${(X^1)}^2$. As explained in Subsection 2.1, this originates in
the fact that in de Sitter space-time there is no positive conserved energy
and implies that ${\cal H}$ can only be used to account for time
evolution in the region where it is non-negative.

\subsection{Solution to the classical equations of motion}

The solution for $X^a$ is the well-known mode sum 
\begin{equation}
 X^a(\tau,\sig) = c_0^a + d_0^a\,\tau + \sum_{n\neq 0}^\infty\,
        i~\frac{c_n^a}{n}~\cos n\sig~e^{-i\,n\tau}\,,
\label{Xa}
\end{equation}
where $c^a_n$ are complex constants of integration (mode amplitudes). Reality
of $X^a$ implies that $c^a_0$ and $d^a_0$ are real and that
$(c^a_n)^\star=c^a_{-n}$. 

The solution for $X^1$ and $X^2$ is more involved. To find it we use separation
of variables $\,X^i(\tau,\sig)=T_i(\tau)\,S_i(\sig)$. This gives 
\begin{align*}
   & \frac{\ddot{T}_1}{T_1}
        = \frac{S^{\prime\prime}_1}{S_1} - m^2\ka^2 = -\la^2_1
   \\[4.5pt]
   & \frac{\ddot{T}_2}{T_2}
        = \frac{S^{\prime\prime}_1}{S_1} + m^2\ka^2 = -\la^2_2
   \,,
\end{align*}
where the dot and prime indicate differentiation with respect to $\tau$ and
$\sig$ respectively. The boundary conditions~(\ref{X1-bc}) and (\ref{X2-bc})
imply that non-trivial solutions are only possible for $\,\la_1\!=\la_2$. We
therefore set $\la\!:=\la_1\!=\la_2$, introduce
\begin{equation}
  \a = \sqrt{\la^2 - m^2\ka^2} \qquad
  \b = \sqrt{\la^2 + m^2\ka^2} 
\label{alpha-beta}
\end{equation}
and distinguish several cases.

\underline{Case 1.}\/: $\la=0$. It is straightforward to see that non-trivial 
solutions only exist if $m\ka$ is an integer. In particular, for 
$m\ka$ an odd integer the solution reads
\begin{align}
    X^1_{\rm o}(\tau,\sig) & =\Big[ \, a_{\rm o} + b_{\rm o}\, \tau
           \sinh\Big(\frac{m\ka\,\pi}{2}\Big) \,\Big]
           \cos(m\ka\,\sig)  \label{zero-1-odd}\\[4.5pt]
    X^2_{\rm o}(\tau,\sig) & =  \frac{B}{m\ka}~b_{\rm o}\>
      \cosh\!\Big[ m\ka\,\Big(\frac{\pi}{2}-\sig\Big)\Big]\,,
    \label{zero-2-odd}
\end{align}
whereas for $m\ka$ an even integer the solution takes the form
\begin{align}
   X^1_{\rm e}(\tau,\sig) & =\bigg[ a_{\rm e} 
       + b_{\rm e}\tau
         \cosh\Big(\frac{m\ka\,\pi}{2} \Big)\,\bigg]
         \cos(m\ka\,\sig) \label{zero-1-even}\\[4.5pt]
   X^2_{\rm e}(\tau,\sig) & =   \frac{B}{m\ka}~b_{\rm e}\>
      \sinh\!\Big[ m\ka\,\Big(\frac{\pi}{2}-\sig\Big)\Big] \,,
   \label{zero-2-even}
\end{align}
with $a_{\rm o},\,b_{\rm o}$ and $a_{\rm e},\,b_{\rm e}$ arbitrary
constants of integration in every instance.

\underline{Case 2.}\/: $\la^2=\pm m^2\ka^2$. This corresponds to either $\a$ or
$\b$ zero and it is very easy to show that the only solution for $X^1$ and
$X^2$ is the trivial one.

\underline{Case 3.}\/: $\la^2\neq 0,\pm m^2\ka^2$. Solving then for $T_i$ and
$S_i$ and imposing the boundary conditions, it follows that the eigenvalues
$\la$ must satisfy the equation 
\begin{equation}
 \big(\la^4 B^4 + \a^2\b^2\big)\,\sin\a\pi\,\sin\b\pi
    - 2\la^2B^2\a\b\,\big (\cos\a\pi\,\cos\b\pi -1\big)=0\,.
\label{eigenvalue}
\end{equation}
Solutions to this equation may occur either because both its terms vanish or
because none of them vanishes but their sum does. We therefore consider two
subcases:

\indent\indent\underline{\emph{Subcase 3.1.}}\/ Both terms in
eq.~(\ref{eigenvalue}) vanish. Since $\a$ and $\b$ are non-zero, we must have
\begin{equation}
    \sin\a\pi \,\sin\b\pi  = \cos\a\pi\,\cos\b\pi -1 =0\,.
\label{eigen-simple}
\end{equation}
It is very easy to see then that the modes for $X^1$ and $X^2$ have the form
\begin{align}
  X^1_{(k,l)}\,(\tau,\sig) & = \frac{i}{\la}\, \Big( 
       a_{\la(k,l)}\frac{\a}{B}~\cos\b\sig 
     + b_{\la(k,l)} \sin\b\sig \Big)\> e^{-i\la\tau} \label{particular-1}\\
    X^2_{(k,l)}\,(\tau,\sig) & = - \Big( 
           b_{\la(k,l)}\>\frac{\b}{\la^2 B}~\cos \a\sig +
          a_{\la(k,l)}\> \sin \a\sig \Big) 
    \> e^{-i\la\tau} \label{particular-2}\,,
\end{align}
where $a_{\la(k,l)}$ and $b_{\la(k,l)}$ are arbitrary constants of
integration. It follows from eqs.~(\ref{eigen-simple}) that $\a$ and $\b$ must
be integers and that their difference must be an even integer. Hence we write
\begin{equation}
  \a = k   \qquad \b=k+2l \,,
\label{case-1-alpha-beta}
\end{equation}
with $k$ and $l$ arbitrary positive integers since $\b\geq\a$ and $\a$
and $\b$ are defined as positive.  With this, equations~(\ref{alpha-beta})
imply
\begin{equation}
    m^2\ka^2\!=2l\,(k+l) > 0 \qquad \la=\pm\>\sqrt{l^2+(l+k)^2}\,.
\label{case-1}
\end{equation}
The first one of these equations states that $m^2\ka^2$ is an even integer. We
thus conclude that for $m^2\ka^2$ an even integer, there are as many modes of
type~(\ref{particular-1})-(\ref{particular-2}) as pairs $(k,l)$ of positive
integers solving the equation $m^2\ka^2=2l(k+l)$, which is clearly a finite
number.

\indent\indent 
\underline{\emph{Subcase 3.2}}\/ We now look at solutions $\la$ to
equation~(\ref{eigenvalue}) such that 
\begin{equation}
  \sin\a\pi \,\sin\b\pi \neq 0\,.
\label{condition}
\end{equation}
In this case the case the modes for $X^1$ and $X^2$ read 
\begin{align}
  X^1_\la(\tau,\sig) & = i  \frac{c_\la}{\la B}~ \Big( \a\, \cos\b\sig
     + \frac{K_\la}{\b}~\sin\b\sig \Big) \, e^{-i\la\tau} 
  \label{X1-sen}\\[4.5pt]
   X^2_\la(\tau,\sig) & = - \Big(\frac{K_\la}{\la^2 B^2}~\cos\a\sig
            + \sin\a\sig \Big) \, e^{-i\la\tau}\,.
  \label{X2-sen}
\end{align}
where $c_\la$ is a arbitrary constant of integration and $K_\la$ is given by
\begin{equation}
  K_\la = \frac{\la^2B^2\, \sin\a\pi + \a\b\,\sin\b\pi}
           {\cos\b\pi - \cos\a\pi}\,.
\label{K}
\end{equation}
Let us study the solutions of equation~(\ref{eigenvalue}) under
condition~(\ref{condition}). Equation~(\ref{eigenvalue}) is an equation in
$\la^2$, so its solutions come in pairs~\hbox{$(\la,-\la)$}.  Solutions with
$\la^2>0$ provide real $\la$ and oscillating degrees of freedom.  By
contrast, solutions with $\la^2<0$ correspond to imaginary $\la$, for which
the $\tau$-exponentials are real.

\indent\indent For $\la^2>0$ and sufficiently large, the left-hand side of the
equation~(\ref{eigenvalue}) can be expanded in powers
of~$x={m^2\ka^2}/{\la^2}\ll 1$, with result
\begin{align}
    {(1+B^2)}^2\, \sin^2\la\pi 
      - x^2\, & \bigg[\>  \frac{\la^2 \pi^2}{4}~{(1-B^2)}^2  
    \label{approximation} \\
    &\!  + (1+B^2)\>\Big( \frac{\la\pi}{8}~ \sin2\la\pi
                     + \sin^2\la\pi \Big)  \bigg] 
      + {\cal O}(x^3) =0\,. \nonumber
\end{align}
The left-hand side is, up to order $x^3$, negative for integer~$\la$ and
positive for non-integer~$\la$. It follows that the left-hand side of
equation~(\ref{eigenvalue}), to which (\ref{approximation}) is an
approximation for large $\la$, must change its sign twice in the vicinity of
every integer $n\!\gg\! |m\ka|$, thus proving the existence of two solutions
around~$n$. These solutions can be found as power series in $\,m\ka/n\,$ by
making for $\la$ in the neighborhood of $n$ the ansatz
\begin{equation*}
  \la_n = n \sum_{k=0}^\infty a_k {\Big(\frac{m\ka}{n}\Big)}^{k} 
  \qquad a_0=1 \,,
\end{equation*}
where the coefficient $a_0$ has been taken equal to 1 since $\la=n$ solves
equation~(\ref{eigenvalue}) to lowest order. Substituting this ansatz in
eq.~(\ref{eigenvalue}) and solving order by order in $m\ka/n$, one obtains two
different sets of solutions for the coefficients $\{a_k\}$, leading to
\begin{equation*}
   \la^{(1,2)}_n = n\, \bigg[ 1 
     \pm \frac{m^2\ka^2}{2\,n^2} ~\frac{1-B^2}{1+B^2} 
     + {\cal O}\bigg(\frac{m^4\ka^4}{n^4}\bigg)\,    \bigg] 
\end{equation*}
This confirms the existence of two real eigenvalues for every large enough
integer~$n$, thus showing that there are infinitely many real solutions with
$|\la|>|m\ka|$. 

\indent\indent By contrast, there is only a finite number of real solutions
with $|\la|<|m\ka|$ and this number depends on the value of $m\ka$. This can
be seen as follows. Assume, without loss of generality, that $m\ka$ is in
between two consecutive integers, so that $\,N\!\leq\!  |m\ka|\! <\! N+1$,
with $N$ a positive integer.  Denote by $N^\prime$ the integer such that
$\,N^\prime\!<\!  \sqrt{2}\, |m\ka|\! \leq\!  N^\prime+1$. Study the sign of
the right-hand side of equation~(\ref{eigenvalue}) as a function of $\b$ by
dividing the interval for $\b$ in subintervals
$[0,1],\,[1,2]\ldots,[N^\prime,N^\prime+1]$.  It is not then very difficult to
prove that
\begin{itemize}
\vspace{-9pt}
\item[(i)] for $N^\prime$ even there are $\,2(N^\prime\!-N+1)\,$ real solutions,
  and 
\vspace{-8.5pt}
\item[(ii)] for $N^\prime$ odd the number of solutions is also
  $2\,(N^\prime\!-N+1)\,$ if
\begin{equation*}
  \frac{2\sqrt{2}}{|m\ka|\,\pi B^2}~ \sin\big(\sqrt{2}|m\ka|\pi\big) 
     + \,\cos\big(\sqrt{2}|m\ka|\pi\big) + 1  >0
\end{equation*}
and $\,2(N^\prime\!-N)\,$ otherwise. 
\vspace{-6pt}
\end{itemize}

\indent\indent We come now to imaginary solutions.  For $\,\la^2\!<0$, with
$\,|\la|\!  >\!m\ka$, the left-hand side of equation~(\ref{eigenvalue}) is
positive definite and never vanishes.  Hence imaginary solutions must have
$|\la|\!<\!m\ka$. Using similar arguments to those employed for real $\la$, it
can be seen that in this case the number of solution for a given \,$m\ka\,$ is
$\,2(N+1)$, with $N$ the integer such that $\,N\!<\!  |m\ka|\! \leq\! N+1$.
We note that imaginary $\la'{\rm s}$ occur due to the different signs with
which $(X^1)^2$ and $X(2)^2$ enter the background metric~(\ref{pp-metric}) and
account for exponential growth of $X^1$ and $X^2$ at $\tau\to\pm\infty$. This
is reminiscent of de Sitter space, for which space expands so fast that light
rays cannot follow.

\indent\indent This analysis shows that there are infinitely many modes of
type~(\ref{X1-sen})-(\ref{X2-sen}), of which a finite number of them have
imaginary $\la$ with $|\la|<|m\ka|$, a finite number have real $\la$ with
$|\la|<|m\ka|$, and infinitely many of them have real $\la$ with
$|\la|>|m\ka|$.  It is important to emphasize that this is so for arbitrary
values of $\,m\ka$, since equation~(\ref{eigenvalue}) and
condition~(\ref{condition}) do not place any limitation on $\,m\ka$. These
modes can also be written in the following way, which will be very useful in
some parts of this paper. The eigenvalue equation~(\ref{eigenvalue}) can be
recast as
\begin{equation*}
   F_+(\la)\,F_-(\la) =0 \,,
\end{equation*}
with $F_\pm(\la)$ functions given by
\begin{equation}
    F_\pm (\la) =\frac{\a\b}{\la^2 B^2}
          - \frac{ (\cos\a\pi \pm 1)\,(\cos\b\pi\mp 1)}
                 {\sin\a\pi\,\sin\b\pi} \,.
\label{F-pm}
\end{equation}
Condition~(\ref{condition}) and the observation that $F_+(\la)$ and $F_-(\la)$
do not have common zeros imply that set of solutions to the eigenvalue
equation~(\ref{eigenvalue}) is the union of the disjoint sets
$\La_+=\{\la_+\}$ and $\La_-=\{\la_-\}$ of solutions of the equations
\begin{equation}
   F_\pm(\la_\pm)=0 \,. 
\label{la-pm}
\end{equation}
It is then a matter of algebra to write $X^1$ and $X^2$ as
\begin{equation}
    X^i_\la(\tau,\sig) = \left\{ \begin{array}{ll}
              X^i_+(\tau,\sig) & ~~ \text{if}~~\la\in\La_+ \\[12pt]          
              X^i_-(\tau,\sig) & ~~ \text{if}~~\la\in\La_- 
                               \end{array}\right. \qquad i=1,2\,,
\label{non-zero}
\end{equation}
with $X^i_\pm$ given by 
\begin{align}
   X^1_\pm(\tau,\sig) & = i\, c_{\la}\> 
     \frac{\a}{\la B}~ 
     \Big( \cos\b\sig + \frac{\sin\b\pi}{\cos\b\pi\mp 1}
                         ~\sin\b\sig \Big) \, e^{-i\la\tau} 
   \label{X1-sol}\\
      X^2_\pm(\tau,\sig) & = - \,c_{\la}\, 
         \Big(\frac{\cos\a\pi\pm 1}{\sin\a\pi}~ \cos\a\sig 
             + \sin\a\sig \Big) \, e^{-i\la\tau}\,.\label{X2-sol}
\end{align}

Putting all cases together, we conclude that the solution for the boundary
problem for $\,X^1,\,X^2\,$ is:
\begin{itemize}
\vspace{-9pt}
\item[(1)] If $m\ka$ is not an integer and its square is not an even
integer, the only modes that occur are those in
eqs.~(\ref{X1-sol})-(\ref{X2-sol}),
corresponding to $\la\in\La_\pm$.
\vspace{-8.5pt}
\item[(2)] If $m\ka$ is not an integer but its square is an even
integer, one has in addition the modes $\,(k,l)\,$
in~(\ref{particular-1})-(\ref{particular-2}).
\vspace{-8.5pt}
\item[(3)] If $m\ka$ is an even integer, there is one additional mode,
$X^1_{\rm e},\,X^2_{\rm e}$
in~(\ref{zero-1-even})-(\ref{zero-2-even}).
\vspace{-8.5pt}
\item[(4)] Finally, if $m\ka$ is an odd integer, the only modes that
occur are those in (1) and $X^1_{\rm o},\,X^2_{\rm o}$
in~(\ref{zero-1-odd})-(\ref{zero-2-odd}).
\vspace{-9pt}
\end{itemize}
We summarize all these situations by writing 
\begin{equation}
  X^i(\tau,\sig) = \sum_{\la\in \La_\pm}\! X^i_\la
       + \dl_{m^2\ka^2,\,{\rm even}}\,\sum_{(k,l)}X^i_{(k,l)}
       + \dl_{m\ka,{\rm even}}\, X^i_{\rm e}
       +  \dl_{m\ka,{\rm odd}}\, X^i_{\rm o} \,.
\label{solution}
\end{equation}

The mode expansions for the momenta $\,P_i,\,P_a\,$ follow from their
expressions~(\ref{Pi})-(\ref{Pa}) in terms of string coordinates and the
mode expansions for the string coordinates.  For the flat $a$-directions it is
trivial to arrive at
\begin{equation*}
   2\pi\a^\prime P_a 
      =  d_0^a + \sum_{n\neq 0}^\infty\, c_n^a~\cos n\sig~e^{-i\,n\tau}\,.
\end{equation*}
For the $i$-directions we have 
\begin{equation}
  P_i(\tau,\sig) = \sum_{\la\in \La_\pm}\! P_{i,\la} 
       + \dl_{m^2\ka^2,{\rm even}}\,\sum_{(k,l)}P_{i,(k,l)}
       + \dl_{m\ka,{\rm even}}\, P_{i,\rm e} 
       + \dl_{m\ka,{\rm odd}}\, P_{i,\rm o} \,,
\label{solution-P}
\end{equation}
where the explicit expressions for the various contributions to the right-hand
side can be found in Appendix A.

\subsection{The string center of mass coordinates and the string total
  momentum}

The string center of mass coordinates
\begin{equation*}
    x^{i,a}_{\rm cm}(\tau) = \frac{1}{\pi}\! \int_0^\pi  d\sig\> 
       X^{i,a}(\tau,\sig) 
\end{equation*}
and the string total momentum
\begin{equation*}
     p_{i,a}(\tau) = \int_0^\pi\!  d\sig\> P_{i,a}(\tau,\sig)
\end{equation*}
are straightforward to calculate from the mode expansions in the previous
subsection. Let us consider for instance the total momentum. For the flat
$a$-directions integration over $d\sig$ gives the standard result
$\,p_a=d^a_0/2\a^\prime$. The $a$-component of the total string momentum is
thus given by one of the string modes in that direction. The situation for the
1 and 2-component is very different.  Indeed, integration over $d\sig$ of the
equations in Appendix A yields
\begin{align}
  p_1(\tau) & = \frac{\dl_{m\ka,{\rm even}}}{\pi\a^\prime}~
        \frac{B^2}{m\ka} b_{\rm e}~\sinh\Big(\frac{mk\pi}{2}\Big) 
        \nonumber\\[3pt]
      & - \frac{m^2\ka^2}{\pi\a^\prime}~\bigg[ 
             \dl_{m^2\ka^2,\,{\rm even}}
             \sum_{{\ssc (k,l)}\atop{\ssc k~{\rm odd}}} 
             \frac{b_{\la(k,l)}}{\b\la^2}~ e^{-i\la\tau} 
          + \sum_{\la\in\La_-} \frac{B\,c_\la}{\b^2}~ 
            \frac{\cos\a\pi-1}{\sin\a\pi} ~e^{-i\la\tau}\,\bigg]
      \label{p1-total}
\end{align}
and 
\begin{align}
  p_2(\tau) & = \frac{\dl_{m\ka,{\rm odd}}}{\pi\a^\prime} \bigg[
          B\,b_{\rm o} \tau \sinh\Big(\frac{mk\pi}{2}\Big) 
        - a_{\rm o} B \bigg] \nonumber \\[3pt]
      & + \frac{im^2\ka^2}{\pi\a^\prime} \bigg[ 
          \dl_{m^2\ka^2,\,{\rm even}}
          \sum_{{\ssc (k,l)}\atop{\ssc k~{\rm odd}}}
                \frac{\la \,a_{\la(k,l)}}{\a}~ e^{-i\la\tau}
          + \sum_{\la\in\La_+} \frac{c_\la}{\la\a}~ e^{-i\la\tau}\bigg] \,.
      \label{p2-total}
\end{align}
The components $p_1$ and $p_2$ receive contributions from all the string modes
in those directions. More importantly, $p_1$ and $p_2$ are not conserved since
their derivatives with respect to $\tau$ do not vanish. This is not a
surprise, for the plane-wave metric~(\ref{pp-metric}) is not invariant under
translations in the 1 and 2-directions. Upon quantization, we therefore do not
expect the eigenvalues of the corresponding operators to play a significant
r\^ole. It is trivial to convince oneself that this collective nature of $p_i$
is also true for the string center of mass coordinates, whose explicit
expression can be trivially obtained through integration over $d\sig$.

\subsection{ Case $\,\boldsymbol{|m\ka|\!\ll\!  1}$}

We finish by considering the regime $\,|m\ka|\!\ll\!  1$. Since $m\ka$ is not
an integer, nor $m^2\ka^2$ is an even integer, the only modes that exist in
this case are those in~(\ref{X1-sen})-(\ref{X2-sen}), or
equivalently~(\ref{X1-sol})-(\ref{X2-sol}). Furthermore, the mode eigenvalues
$\la$ can be explicitly found as formal power series in $m\ka$ by making the
ansatz $\,\la=\sum_{0}^\infty b_k (m\ka)^k$ and solving
equation~(\ref{eigenvalue}) for the coefficients $b_k$ order by order.
Proceeding in this way we obtain: 

\emph{(i) Imaginary
  eigenvalues.} As already mentioned, they have $|\la|\!<|m\ka|$. The algebra
shows that there are only two of them, $\,\La^{\rm I}=\{\pm i\la^{\rm I}$\},
given by
\begin{equation}
  \la^{\rm I}  = \frac{m\ka}{\sqrt{1+B^2}}~ \bigg[ 
       1 + \frac{(m\ka)^2}{12}~\frac{\pi^2\,B^2}{1+B^2}
         + \frac{(m\ka)^4}{1440}~ \frac{\pi^4\,B^2\,(5\,B^2-24)}{{(1+B^2)}^2}
         + {\cal O}\big(m^6\ka^6\big)\, \bigg]\,.
\label{eigen-left-im}
\end{equation}
In terms of equations~(\ref{la-pm}), they happen to solve $F_-(\la)=0$, thus
belong to $\La_-$.

\emph{(ii) Real eigenvalues with $|\la|\!<|m\ka|$.} There are also two of
them,~$\,\La^{\rm R}\!= \{\pm\la^{\rm R}\}$, where
\begin{equation}
    \la^{\rm R}  = \frac{m\ka}{\sqrt{1+B^2}}~\bigg[ 1 
     - \frac{(m\ka)^2}{12}~\frac{\pi^2\,B^2}{1+B^2}  
     + \frac{(m\ka)^4}{1440}~ \frac{\pi^4\,B^2\,(5\,B^2-24)}{{(1+B^2)}^2} 
     +{\cal O}\big(m^6\ka^6\big)\, \bigg] \,.
\label{eigen-left-re}
\end{equation}
They are now solutions of $F_+(\la)=0$, thus are in $\La_+$.

\emph{(iii) Real eigenvalues with $|\la|\!>|m\ka|$.} They read
\begin{equation}
  \left\{\begin{matrix} \la_n \\ \tilde{\la}_n \end{matrix} \right\} 
    = n \, \bigg[ 1\,  \pm
       \,\frac{m^2\ka^2}{2\,n^2} ~\frac{1-B^2}{1+B^2} 
       - \frac{m^4\ka^4}{8\,n^4}~ \frac{B^4-6B^2+1}{(1+B^2)^2} 
       + {\cal O}\big(m^6\ka^6\big)\,\bigg] \,,
\label{eigen-right}
\end{equation}
where $n$ is a non-zero integer and the $\,+/-\,$ signs on the right-hand side
correspond to $\,\la_n/\tilde{\la}_n\,$ on the left side. We will use the
notation $\La:=\{\la_n\}$ and $\tilde{\La}:=\{\tilde{\la}_n\}$. These
eigenvalues can be reorganized in terms of solutions of
equations~(\ref{la-pm}) as
\begin{equation}
   \la_+^{(n)} = \left\{ \begin{matrix} \la_n & {\rm if}~n~{\rm even} \\[3pt]
                      \tilde{\la}_n & {\rm if}~n~{\rm odd}
                     \end{matrix}   \right. \qquad
   \la_-^{(n)} = \left\{ \begin{matrix} \tilde{\la}_n & {\rm if}~n~{\rm even} 
                       \\[3pt]
                      \la_n & {\rm if}~n~{\rm odd} \end{matrix} \right. \,.
\label{reorganized}
\end{equation}

It is instructive to compare the mode eigenvalues with those for the open
string in flat space-time and zero antisymmetric field, i.e. with $m=B=0$. In
that case, $X^1$ and $X^2$ have the same expansion as in~(\ref{Xa}) and the
mode eigenvalues are the integers. The flat zero mode $\,\la_{\rm flat}\!=0$
has multiplicity four in the 1,2-directions, for there are four arbitrary
constants of integration, which in our notation would be denoted
$\,c^1_0,\,c^2_0,\,d^1_0,\,d^2_0$. Every pair $\,(n,-n)\,$ of non-zero flat
modes is also 4-degenerate in these directions, for in each direction there
are two complex coefficients $c^i_{-n}$ and $c^i_n$ and one complex constraint
${(c^i)}^\star_n=c^i_{-n}$. If $m$ and $B$ are switched on, the flat zero mode
unfolds into two non-zero imaginary modes $\,(i \la^{\rm I},-i\la^{\rm I})\,$
and two non-zero real modes $\,(\la^{\rm R},-\la^{\rm R})$, and every pair of
flat modes $(n,-n)$ unfolds into four modes
$(\la_n,\tilde{\la}_n,\la_{-n},\tilde{\la}_{-n})$.  Whereas in Minkowski
space-time, the string center of mass and string total momentum are
independent degrees of freedom associated to the 4-degenerate zero mode, in
our plane-wave background they are collective quantities.

\section{Quantization}

There is a discussion in the literature for $m=0$ as for how to quantize the
open string with non-trivial boundary conditions like those in~(\ref{X1-bc})
and~(\ref{X2-bc}). It seems to be a widespread believe that these boundary
conditions impeach the use of canonical quantization. In fact, for $m=0$,
Dirac quantization, with the boundary conditions regarded as constraints, has
been used as an alternative. The problem that arises then is whether the
boundary conditions should be regarded as first or second class, and this is
not a trivial choice for they lead to different
results~\cite{Ardalan}~\cite{Chu-Ho-2}.

We will use plain canonical quantization and show that there is nothing wrong
with it. Our approach consists of two steps. In the first one we compute the
symplectic form in terms of the modes. This is straightforward, since the
action is first order in time derivatives and it is well-known how to proceed
in these cases~\cite{Witten-CMP}~\cite{Faddeev-Jackiw}. The resulting
symplectic form will be non-singular, so it has an inverse. Its inverse
defines, upon standard canonical quantization~\cite{Witten-CMP}, the
commutation relations for the quantum theory. We emphasize that the
calculation of the symplectic form may be involved but ,as pointed out in
refs.~\cite{Witten-CMP}~\cite{Faddeev-Jackiw}, as far as it is non-singular
there is nothing wrong with canonical quantization and there is no need to
introduce constraints of any type. It is also worth noting in this respect
that the boundary conditions have already been taken into account in solving
the classical equations of motions, so one would na\"ively expect the
symplectic form to already account for them. We will see that this
quantization method is consistent with the equal-time commutation relations
\begin{equation}
   \big[\,X^i(\tau,\sig),P_j(\tau,\sig^\prime)\,\big] 
               = i\,\dl^i_{\>j}\,\dl(\sig-\sig^\prime)\,.
\label{canonical}
\end{equation}
In Section 5 we will explicitly construct the Fock-Hilbert space for the
theory and find the hamiltonian spectrum.

\subsection{Symplectic form  and canonical quantization}

The symplectic form
\begin{equation*}
  \Omega =  \int_0^\pi\! d\sig \>\big( \dr P_i\wedge\dr X^i 
                 + \dr P_a\wedge \dr X^a \big) \,.
\end{equation*}
is the sum of two contributions, which we will call $\,\Omega_{pp}$ and
$\Omega_{\rm flat}$. They respectively arise from the modes in the
$i$-directions and the flat $a$-directions.  Since they do not mix, the
symplectic form can be studied by separately looking at each one of these
two sectors.

Let us first look at $\Omega_{pp}$. Recalling the mode expansions for $X^a$
and $P_a$, one easily arrives at
\begin{equation}
   \Omega_{\rm flat} = \int_0^\pi\! d\sig \> \dr P_a\wedge \dr X^a  
            = \frac{1}{2\a^\prime}  \sum_{a=3}^{D-2} \> \Big( 
      \dr d^a_0 \wedge \dr c^a_0 
    - \sum_{n\neq0}\frac{i}{2n}~\dr c^a_n \wedge \dr c^a_{-n}\Big) \,.
\label{symplectic-flat}
\end{equation}
This can be written as
\begin{equation}
   \Omega_{\rm flat} = \frac{1}{2}~ \Omega_{MM^\prime}\; 
            \dr A_M\!\wedge \dr A_{M^\prime}\,
\label{quantization-1}
\end{equation}
where $\{A_M\}=\{d^a_0,c_0^a,c_n^a\}$ and a summation over indices $M=(a,n)$
and $M^\prime=(a^\prime,n^\prime)\,$ is understood. The form $\Omega_{\rm
  flat}$ is non-singular and can be inverted. Upon quantization, the
amplitudes $\{A_M\}$ become operators with commutation relations given by the
inverse of $\Omega$ as
\begin{equation}
    [A_M,A_{M^\prime}]= i \>\big(\Omega^{-1}\big)_{MM^\prime}\,.
\label{quantization-2}
\end{equation}
This yields the standard commutation relations
\begin{equation*}
  [c^a_0,d^b_0] = 2i\a^\prime \dl^{ab} \qquad 
  [c^a_n,c^b_m] = 2\a^\prime n\, \dl^{ab}\,\dl_{n+m,0}\,.
\label{commutators-ab}
\end{equation*}
Reality of the field operators $X^a$ imply that $c^a_0$ and $d^a_0$ are
hermitean and that $c^a_{-n}=(c^a_n)^\dagger$. So far, this is the same
analysis as for Minkowski spacetime.

To compute $\Omega_{pp}$ it is most convenient to use equation~(\ref{Pi}) and
write $P_i$ in terms of derivatives of $X^i$ with respect to $\tau$ and
$\sigma$. This gives
\begin{equation*}
  \Omega_{pp}  = \int_0^\pi\! d\sig \>\dr P_i\wedge\dr X^i 
  = \tilde{\Omega}_{pp} + \bar\Omega_{pp} \,, 
\end{equation*}
where $\tilde\Omega_{pp}$ and $\bar\Omega_{pp}$ read
\begin{equation}
  \tilde\Omega_{pp} = \frac{1}{2\pi\a^\prime} \int_0^\pi\!d\sig \,
                   \>\dr(\pa_\tau X^i)\! \wedge \dr X^i 
\label{bulk}
\end{equation}
and 
\begin{equation}
   \bar\Omega_{pp} =  \frac{B}{2\pi\a^\prime}~\> 
    \dr X^1\! \wedge \dr X^2\, \bigg\vert_{\sig=0}^{\sig=\pi} \,.
\label{boundary}
\end{equation}
As compared to the flat $a$-directions, for which $2\pi\a^\prime P_a=\pa_\tau
X^a$, the nontrivial boundary conditions not only modify the modes but also
add a boundary term $\bar\Omega_{pp}$ to the symplectic form.  Computation of
the boundary piece $ \bar\Omega_{pp}$ is straightforward. To calculate
$\tilde\Omega_{pp}$, we use the mode expansion~(\ref{solution}), integrate
over $d\sig$, rearrange the mode sums and employ that the eigenvalues
$\la\in\La_\pm$ are solutions of equations~(\ref{la-pm}). After some very
long, but also very straightforward algebra, we obtain that
\begin{equation}
  \Omega_{pp} = \Omega_{\La_\pm} 
     + \dl_{m^2\ka^2,\rm even}~ \Omega_{\{(k,l)\}}  
     + \dl_{m\ka,\rm even}~ \Omega_{\rm e} 
     + \dl_{m\ka,\rm odd}~ \Omega_{\rm o} \,.
\label{symplectic}
\end{equation}
The various contributions in this equation are given by
\begin{eqnarray}
   &{\ds
  \Omega_{\La_\pm} =  \frac{i}{2\pi\a^\prime}\sum_{\la\in\La_\pm} f(\la)~
        \dr c_\la \wedge \dr c_{-\la} }\\[3pt]
   &{\ds  \Omega_{\{(k,l)\}} =- \frac{i}{4\a^\prime B}\, \sum_{(k,l)}\,
     \big[ \,f_a(\la)~ \dr a_{\la(k,l)} \wedge \dr a_{-\la(k,l)}
          + f_b(\la)~ \dr b_{\la(k,l)} \wedge \dr b_{-\la(k,l)}\, \big] 
    }\\[3pt]
   &{\ds \Omega_{\rm e}  = - \frac{1}{4\a^\prime}~ 
               \cosh\Big(\frac{m\ka\pi}{2}\Big)~
               \dr a_{\rm e} \wedge \dr b_{\rm e} }\\[4.5pt]
   &{\ds \Omega_{\rm o}  = -\frac{1}{4\a^\prime}~
                \sinh\Big(\frac{m\ka\pi}{2}\Big)~    
                \dr a_{\rm o} \wedge \dr b_{\rm o} }\,,
\end{eqnarray}
where $\,f(\la),\,f_a(\la)$\, and $\,f_b(\la)\,$ read
\begin{eqnarray}
   &{\ds f(\la) = -\,\frac{\la\a\,(\cos\a\pi \pm 1)}{\sin\a\pi}\>
     \bigg[\,\frac{2\,(m\ka)^4}{\la^2\a^2\b^2} \pm 
     \frac{\pi}{\a\sin\a\pi} \mp \frac{\pi}{\b\sin\b\pi}\,\bigg] }&
    \label{f-la} \\[3pt]
   &{\ds f_a(\la) = \frac{\la}{B^2}~\Big( 1+B^2 
               - \frac{m^2\ka^2}{\la^2} \Big)  }&\label{f-la-a}\\[3pt]
   &{\ds f_b(\la)  = \frac{1}{\la B^2}~\Big( 1+B^2 
               + \frac{m^2\ka^2}{\la^2} \Big)\,. }& \label{f-la-b}\
\end{eqnarray}
In accordance with the notation that we are using, the double signs $\pm$ on
the right of the equation for $\,f(\la)\,$ apply, respectively, to the
eigenvalues $\la_\pm$ solving the equations~~(\ref{la-pm}).

We make at this point two comments concerning the computation of
$\Omega_{pp}$. The first one is that the only non-zero components
$\Omega_{MM^\prime}$ of the symplectic form have $M+M^\prime=0$, where $M$
labels all the existing mode $\{A_M\}=\{a_{\rm o},b_{\rm o},a_{\rm e},b_{\rm
  e}, a_{\la(k,l)}, b_{\la(k,l)}, c_\la\}$. Some authors call this
orthogonality of modes. Note in particular that there is not any mixing of the
modes for $\,m^2\ka^2\!={\rm even}$, \,$m\ka={\rm even}\,$ and $\,m\ka={\rm
  odd}$ among themselves, nor with modes $\la\!\in\!\La_\pm$.  The second
comment is to emphasize that the result above for $\Omega_{pp}$ follows
straightforwardly from eqs.~(\ref{bulk})-(\ref{boundary}) after plain
integration over $d\sig$, without any assumption whatsoever.

The form $\Omega_{pp}$ is non-singular and has an inverse $\Omega_{pp}^{-1}$.
Canonical quantization is then straightforward. The amplitudes $\,\{A_M\}\,$
become operators. Hermiticity of $X^i$ implies that~$a_{\rm o}$, $b_{\rm o}$,
$a_{\rm e}$, $b_{\rm e}$ and $c_\la$ ($\la\!\in\!\La_\pm$ imaginary) are
hermitean and that
\begin{equation*}
    a_{\la(k,l)}^\dagger\! = a_{-\la(k,l)} \qquad 
    b_{\la(k,l)}^\dagger\! = b_{-\la(k,l)} \qquad 
    c_\la^\dagger\! = c_{-\la}~(\la\!\in\!\La_\pm~{\rm real})\,.
\end{equation*}
The commutation rules are obtained from the inverse of $\Omega_{pp}$ as
in~(\ref{quantization-1})-(\ref{quantization-2}), the only non-trivial
commutation relations being
\begin{eqnarray}
    &{\ds \big[\,c_\la,c^\dagger_\la\,\big] = -\,\frac{\pi\a^\prime}{f(\la)} 
     }& \label{commutators-12}\\
    &{\ds \big[\,a_{\la(k,l)},a^\dagger_{\la(k,l)}\,\big] 
             = \frac{2\a^\prime}{f_a(\la)} \qquad 
          \big[\,b_{\la(k,l)},b^\dagger_{\la(k,l)}\,\big] 
             = \frac{2\a^\prime}{f_b(\la)} } & 
    \label{comm-ab}\\[4.5pt]
   &{\ds \big[\,a_{\rm e},b_{\rm e}\,\big] = -\,4i\a^\prime\,   
         {\rm cosech}\Big(\frac{m\ka\pi}{2}\Big)} & 
    \label{comm-even}\\[4.5pt]
   &{\ds \big[\,a_{\rm o},b_{\rm o}\,\big] = -\,4i\a^\prime\, 
         {\rm sech}\Big(\frac{m\ka\pi}{2}\Big)\,. } & \label{comm-odd}
\end{eqnarray}
We note that $f(\la)$ is real for $\la$ real and imaginary for $\la$
imaginary. The space of states on which these operators act and their action
is given in the Section 5. Let us move on to study the consistency of this
quantization with the canonical commutation relations (\ref{canonical}).

\subsection{Canonical commutation relations}

The commutator $\,[X^i(\tau,\sig),P_j(\tau,\sig^\prime)]\,$ can be computed by
replacing $X^i$ and $P_j$ with their mode expansions and using the
relations~(\ref{commutators-12})-(\ref{comm-odd}) for the mode operators in
them. In doing so, the $\tau$-dependence of the commutator is removed and a
mode sum is left. This sum involves in particular an infinite sum over mode
eigenvalues $\la\!\in\!\La_\pm$ whose terms are products of sines and cosines
at $\a\sig,\,\b\sig,\,\a\sig^\prime,\,\b\sig^\prime$ with complicated
coefficients involving the function $f(\la)$. We do not see a way to perform
this sum in closed form and obtain a compact expression for the commutator. We
will instead expand the commutator in powers of $m\ka$ and perform the mode
sums order by order in $m\ka$. We do this in the sequel.

If $|m\ka|\!\ll\!  1$, the only modes that exist are those in
eqs.~(\ref{X1-sol})-(\ref{X2-sol}). We recall from Subsection 2.4 that in this
case the mode eigenvalues are given by $\,\La_{\rm I}\!=\!\{\pm i\la^{\rm
  I}\}$, $\,\La_{\rm R}\!=\!\{\pm\la^{\rm R}\}$, $\,\La\!=\!\{\la_n\}$ and
$\,\tilde{\La}\!=\!\{\tilde{\la}_n\}\,$ in
eqs.~(\ref{eigen-left-im})-(\ref{eigen-right}), with $n=\pm1, \pm2, \ldots$ We
denote by $\,\{ c_\pm^{\rm I}\}$, $\{c_\pm^{\rm R}\}$, $\{c_{n}\}$ and
$\{\tilde{c}_{n} \}\,$ the corresponding annihilation and creation operators,
for which hermiticity of the string position operators implies
\begin{equation*}
  (c_\pm^{\rm  I})^\dagger=c_\pm^{\rm  I} \quad 
  (c_+^{\rm  R})^\dagger=c_-^{\rm R} \quad 
  (c_{n})^\dagger= c_{-n} \quad 
  (\tilde{c}_{n})^\dagger=\tilde{c}_{-n}\,.
\end{equation*}
Expanding the right-hand-side of eq.~(\ref{commutators-12}) in powers of
$m\ka$, we obtain the following commutations relations for them:
\begin{align}
   [c^{\rm  I}_+,c^{\rm  I}_-] & = -\,
         \frac{i\a^\prime B^2}{\,2\,(2+B^2)\,(1+B^2)^{1/2}}~ \frac{1}{m\ka}~
     \bigg[\, 1 + \frac{\,\pi^2 B^2\, (m\ka)^2\,}{6\,(2+B^2)} 
       +\,  {\cal O}\big(m^4\ka^4\big) \,\bigg] \label{comm-I} \\[6pt] 
   [c^{\rm  R}_+,c^{\rm  R}_-] & = -\, 
            \frac{\a^\prime \pi^2 B^2}{\,8\,{(1+B^2)}^{3/2}}~\,m\ka~
    \bigg[\, 1 +\, \frac{\,\pi^2 (1-B^2)\, (m\ka)^2\,}{6\,(1+B^2)}
         +\,  {\cal O}\big(m^4\ka^4\big) \,\bigg]  \label{comm-R} \\[6pt]
   [c_n,c_k] & = \frac{\a^\prime \pi^2 B^4}{\,4n^3\,{(1+B^2)}^3}~
    (m\ka)^4~
    \bigg[\, 1 -\, \frac{\,(3-5B^2)\, (m\ka)^2\,}{2n^2\,(1+B^2)}  
            +\,  {\cal O}\big(m^4\ka^4\big) \,\bigg]\>  \dl_{n+k,0} 
   \label{comm-n} \\[6pt]
   [\tilde{c}_n,\tilde{c}_k] & = \frac{\a^\prime B^2}{\,n\,(1+B^2)\,}~ 
    \bigg[\, 1 + \,\frac{3\,(m\ka)^2}{\,2n^2(1+B^2)\,}
            +\, {\cal O}\big(m^4\ka^4\big)\, \bigg]\>  \dl_{n+k,0}\,,
   \label{comm-n-tilde} 
\end{align}
all other commutators being zero. The commutator
$\,[X^i(\tau,\sig),P_j(\tau,\sig^\prime)]\,$ can then be written as a sum
\begin{equation*}
   [X^i(\tau,\sig),P_j(\tau,\sig^\prime)] = \sum_{\om=I,R,n,\tilde{n}} 
    C^{\,i}{}_{\!j}(\om;\sig,\sig^\prime)
\end{equation*}
of four contributions $\,C^i{}_{\!j}(\om;\sig,\sig^\prime)$ arising from the
four sets in which the modes have been organized. Each one of these
contributions is a power series in $m\ka$, depends on $\sig$ and $\sig^\prime$
and can be computed with relative ease order by order. To illustrate this, let
us take as an example $i=j=2$. After some algebra we obtain
\begin{align}
  C^{\,2}{}_{\!2}({\rm I}; \sig,\sig^\prime) & = -\,
  \frac{\,i\a^\prime B^2\, (m\ka)^2\,}{2\,(1+B^2)}~\Big( 
      \frac{\pi^2}{2} -\pi\sig -\pi\sig^\prime + 2\sig\sig^\prime \Big)
       +  {\cal O}\big(m^4\ka^4\big)  \label{imaginary} \\[3pt]
  C^{\,2}{}_{\!2}({\rm R}; \sig,\sig^\prime) & = i\a^\prime  
      - \frac{\,i\a^\prime B^2\,(m\ka)^2}{2\,(1+B^2)}~ \Big( \frac{\pi^2}{6} 
        + \pi\sig -\sig^2 -\pi \sig^\prime +{\sig^\prime}^2\Big)   
      +\, {\cal O}\big(m^4\ka^4\big) \label{real}\\[3pt]
  C^{\,2}{}_{\!2}(\La;\sig,\sig^\prime) & = 2i\a^\prime 
      \sum_{n=1}^\infty \cos n\sig^\prime\> \bigg[ \cos n\sig 
   + \frac{\,B^2\, (m\ka)^2\,}{1+B^2}~ \Big( \frac{\cos n\sig}{n^2}
        + \frac{\sig\sin n\sig}{n} - 
        \frac{\pi}{2}\> \frac{\sin n\sig}{n} \Big) \bigg] \notag \\[3pt]
   & + \, {\cal O}\big(m^4\ka^4\big) \label{Lambda} \\[3pt]
   C^{\,2}{}_{\!2}(\tilde{\La};\sig,\sig^\prime) & =-\> 
       \frac{\,i\a^\prime B^2\,(m\ka)^2\,}{1+B^2} \sum_{n=1}^\infty 
           ~ (2\sig^\prime-\pi)~\frac{\sin n\sig \,\cos n\sig^\prime}{n} 
       + \, {\cal O}\big(m^4\ka^4\big)\,. \label{tilde-Lambda}
\end{align}
It follows from inspection of these formuli that only $C^{\,2}{}_{\!2}({\rm
  R})$ and $ C^{\,2}{}_{\!2}(\La)$ carry contributions of order zero in $m\ka$.
These are easily summed by recalling that, for functions defined on $[0,\pi]$
with vanishing derivatives at the boundary, Dirac's delta function has the
representation
\begin{equation*}
   \pi\, \dl(\sig-\sig^\prime) =  1 
       + 2  \sum_{n=1}^\infty \cos n\sig \, \cos n\sig^\prime \,.
\end{equation*}
Hence
\begin{equation*}
    [X^2(\tau,\sig),P_2(\tau,\sig^\prime)]_0 
       =  i\a^\prime \dl(\sig-\sig^\prime)\,,
\end{equation*}
where the subscript 0 refers to the order in $m\ka$. To sum the order-two in
$m\ka$ contributions, it is convenient to introduce variables
$\sig_\pm=\sig\pm\sig^\prime$, which take values $\,\sig_-\!\in[-\pi,\pi]\,$
and $\,\sig_+\!\in[0,2\pi]$. In terms of these, we have
\begin{equation*}
  \Big[ C^{\,2}{}_{\! 2}(\La) + C^{\,2}{}_{\! 2}(\tilde{\La}) \Big]_2 
  = \,\frac{\,i\a^\prime B^2\,(m\ka)^2\,}{2\,(1+B^2)}~
          \big[\, F_2(\sig_-) + F_2(\sig_+) 
      + \sig_-  F_1(\sig_-)  + \sig_- F_1(\sig_+) \,\big] \,,
\end{equation*}
where $F_1$ and $F_2$ stand for the Fourier series
\begin{align}
  F_1(\sig_-):= 2 \sum_{n=1}^\infty \frac{\sin n\sig_-}{n} & =  \left\{ 
   \begin{array}{ll}  {\ds \frac{\pi|\sig_-|}{\sig_-}} - \sig_-  
                               & {\rm if} ~~ 0<|\sig_-|<\pi \\[9pt]
                       0  & {\rm if } ~~ \sig_-=0,\pm \pi \end{array} 
  \right. \label{F1} \\[6pt]
   F_1(\sig_+) := 2 \sum_{n=1}^\infty \frac{\sin n\sig_+}{n} & = 
    \left\{ \begin{array}{ll} \pi -\sig_+  & {\rm if} ~~ 0<\sig_+<2\pi 
      \\[6pt]  0  & {\rm if } ~~ \sig_+=0,2\pi \end{array} \right.\,.
  \label{F2} \\[6pt]
  F_2(\sig_-) := 2 \sum_{n=1}^\infty \frac{\cos n\sig_-}{n^2} & = 
        \frac{\sig^2_-}{2} -\pi\,|\sig_-| + \frac{\pi^2}{3} 
  \label{F3} \\[6pt]
   F_2(\sig_+):= 2 \sum _{n=1}^\infty \frac{\cos n\sig_+}{n^2} & = 
        \frac{\sig_+^2}{2} -\pi\, \sig_+ +\frac{\pi^2}{3} 
  \label{F4} \,.
\end{align}
Putting together all contributions of order two in
eqs.~(\ref{imaginary})-(\ref{tilde-Lambda}), we obtain 
\begin{equation*}
  \big[X^2(\tau,\sig),P_2(\tau,\sig^\prime)\big]_2 =0 \, ,
\end{equation*}
in agreement with~(\ref{canonical}). Proceeding in the same way, it is
straightforward to see that the commutation relations in
(\ref{canonical}) also hold for other values of $i$ and $j$, so we can write
\begin{equation*}
    [X^i(\tau,\sig),P_k(\tau,\sig^\prime)] = 
       i\a^\prime\,\dl^i{}_{\!j}\, \dl(\sig-\sig^\prime) 
       +\,{\cal O}\big(m^4\ka^4)\,.  
\end{equation*}
This proves the consistency of the quantization procedure used here with
equal-time canonical commutation relations, at least up to order $m^4\ka^4$.

We find quite surprising the asymmetric r\^ole that each type of mode plays in
this analysis, yet all combine to produce the desired result. It is also worth
noting that $\,C^i{}_{\!j}(\La)\,$ will involve to any order in $m\ka$
polynomials in $\sig_\pm$ multiplied with convergent Fourier series of
$\sig_\pm$, thus becoming a question of algebra force to go to higher orders
in $m\ka$. It is by now clear that canonical quantization works and that it
does because the symplectic form is non-singular.

\section{Non-commutative wave fronts}

The plane-wave metric~(\ref{pp-metric}) foliates spacetime by null surfaces
$X^+={\rm const}$. We show next that these spaces are non-commutative. The
commutator $\,\big[X^1(\tau,\sig), X^2(\tau,\sig^\prime)\big]\,$ can be
computed by replacing $\,X^1\,$ and $\,X^2\,$ with their mode expansions and
using the commutation relations~(\ref{commutators-12})-(\ref{comm-odd}) for
the mode operators.  This results in
 \begin{align}
   \big[X^1(\tau,\sig),&X^2(\tau,\sig^\prime)\big] = i\,\Big[
   \Theta_{\La_\pm}(\sig,\sig^\prime)
    \nonumber \\[3pt]
    & + \dl_{m^2\ka^2,\rm even}~\Theta_{\{(k,l)\}}(\sig,\sig^\prime) 
      + \dl_{m\ka,\rm even}~\Theta_{\rm e}(\sig,\sig^\prime)
            +  \dl_{m\ka,\rm odd}~\Theta_{\rm o}(\sig,\sig^\prime)\,\Big]
\label{X1X2-general}
\end{align}
where the contribution $\Theta_{\La_\pm}(\sig,\sig^\prime)$ is given by
\begin{align}
   \Theta_{\La_\pm}(\sig,\sig^\prime) = \frac{1}{2B} 
    & \sum_{\la\in\La_\pm} \frac{\a}{\la \,f(\la)} \>\Big( \cos\b\sig +
    \frac{\sin\b\pi}{\cos\b\pi\mp 1} ~\sin\b\sig \Big)
    \nonumber \\[6pt]
    & \,{\scriptstyle \times} \,  \>
    \Big(\frac{\cos\a\pi\pm 1}{\sin\a\pi}~ \cos\a\sig ^\prime +
    \sin\a\sig^\prime \Big)   \label{Th-general} 
\end{align}
and $\Theta_{\{(k,l)\}}(\sig,\sig^\prime),\>\Theta_{\rm
  e}(\sig,\sig^\prime)\,$ and $\,\Theta_{\rm o}(\sig,\sig^\prime)\,$ read
\begin{eqnarray}
   &{\ds \Theta_{\{(k,l)\}}(\sig,\sig^\prime) = - 4\a^\prime B \sum_{(k,l)}
    \bigg[ \frac{\a\cos\b\sig\,\sin\a\sig^\prime}{\la^2\,(1+B^2) - m^2\ka^2}  
         + \frac{\b\sin\b\sig\,\cos\a\sig^\prime}{\la^2\,(1+B^2) + m^2\ka^2} 
         \bigg] } \label{Th-square} \\[3pt]
   &{\ds \Theta_{\rm e}(\sig,\sig^\prime) = \frac{4\a^\prime B}{m\ka}~ 
       {\rm cosech} \Big(\frac{m\ka\pi}{2} \Big)\,\cos(m\ka\sig)\,
       \sinh \Big[ m\ka\,\Big(\frac{\pi}{2}-\sig^\prime\Big)\,\Big] }& 
    \label{Th-even} \\[3pt]
   &{\ds \Theta_{\rm o}(\sig,\sig^\prime) = \frac{4\a^\prime B}{m\ka}~
       {\rm sech} \Big(\frac{m\ka\pi}{2} \Big)\,\cos(m\ka\sig)\,
       \cosh \Big[ m\ka\,\Big(\frac{\pi}{2}-\sig^\prime\Big)\,\Big] \,.} &
   \label{Th-odd}
\end{eqnarray}
We recall that the sum in $\Theta_{\{(k,l)\}}(\sig,\sig^\prime)\,$ is over the
finite number of solutions $\,(k,l)\,$ of equation~(\ref{case-1}) and that
$\a$ and $\b$ in this sum are as in~(\ref{case-1-alpha-beta}), so the
contributions~(\ref{Th-square})(\ref{Th-odd}) do not pose any problems.

The most complicated piece to understand is the
contribution~$\,\Theta_{\La_\pm}(\sig,\sig^\prime)$.  We may proceed as in
Section 3 and consider $|m\ka|\!\ll\! 1$. In this case only
$\Theta_{\La_\pm}(\sig,\sig^\prime)$ contributes to the commutator
$[X^1,X^2]$.  Expanding the right-hand side of equation~(\ref{Th-general}) in
powers of $m\ka$, the sum over modes can then be performed order by order in
$\,m\ka$, so that $\Theta(\sig,\sig^\prime)$ becomes a power series
\begin{equation*}
  \Theta(\sig,\sig^\prime) =
        \sum_{k=0}^\infty \Theta_{2k}(\sig,\sig^\prime)
  \>(m\ka)^{2k} 
\end{equation*}
whose coefficients are explicit functions of $\sig$ and $\sig^\prime$. The
first two terms of this series are calculated in Appendix B. We exhibit
here the result. At the string endpoints we obtain
\begin{equation}
   \Theta(0,0) =  - \,\Theta(\pi,\pi) 
    = \frac{\a^\prime\pi B}{1+B^2}~
     \bigg[ 1 + \frac{\pi^2\,(m\ka)^2}{6\,(1+B^2)} 
          + \,{\cal O}\big(m^4\ka^4\big) \bigg]\,,
\label{Th-ends}
\end{equation}
whereas at $\,\sig+\sig^\prime\neq 0,2\pi\,$ we have
\begin{align} 
   \Theta(\sig,\sig^\prime) & 
  = \frac{\a^\prime B\,(m\ka)^2}{\,{(1+B^2)}^2\,}~
    \bigg\{ B^2 \Big[\! - \frac{\sig}{6}\,\big(\sig^2-3{\sig^\prime}^2\big) 
         + \frac{\pi}{4}\,\big(\sig^2- {\sig^\prime}^2- 2\sig\sig^\prime\big)
         - \frac{\pi}{12}\,\big(\sig-3\sig^\prime\big) \Big] 
         \nonumber \\[3pt]
  & - \frac{\sig}{12}\,\big(7\sig^2+9{\sig^\prime}^2\big) 
    + \frac{\pi}{8}\,\big(7\sig^2+3{\sig^\prime}^2+6\sig\sig^\prime\big)
    - \frac{\pi^2}{4}\,\big(3\sig+\sig^\prime\big) + \frac{\pi^3}{6} 
    \nonumber \\[3pt]
  & + \frac{\pi}{8}\,|\sig-\sig^\prime|\,
    \Big[\, 2B^2\, \big(\sig^2+\sig^\prime-\pi\big)
         + 5\sig-\sig^\prime-2\pi \Big] \,\bigg\}
         +\,{\cal O}\big(m^4\ka^4\big)\,.
\label{Th-inner}
\end{align}
The limit $m\to 0$ is smooth and reproduces the results in the literature. In
fact, as $m\to 0$, that is, as Minkowski spacetime s approached, only the
first term in~(\ref{Th-ends}) survives and the results in ref.~\cite{Chu-Ho-1}
are recovered. For $m\neq 0$, two novelties are found: non-commutativity at
the string endpoints receives \emph{m}-dependent corrections, and
non-commutativity occurs for arbitrary values of $\sig$ and $\sig^\prime$, so
that it extends all along the string. Even for $\sig=\sig^\prime\neq0,\pi$
non-commutativity pervades, since in that case
\begin{equation*}
   \Theta(\sig,\sig) = \frac{\a^\prime B\,(m\ka)^2}{6\,{(1+B^2)}^2}~
        (2\sig-\pi)\, \Big[\, B^2 \sig\,(\sig-\pi) 
        - \big(2\sig-\pi\big)^2 \Big] +\,{\cal O}\big(m^4\ka^4\big)\neq 0\,.
\end{equation*}
At the string midpoint $\sig\!=\!\sig^\prime\!=\!\pi/2\,$ one has
$\Theta=0$, not only for $\,m^2\ka^2\!\ll\!1\,$ but also for arbitrary $m\ka$
since the right-hand side of~(\ref{Th-general}) vanishes. 
Note also that commutativity is recovered as $B\to 0$.

The results in this Section may be viewed from two perspectives.  The first
one is to assume a constant background field $B_{12}=B$ and that the string
endpoints move freely, except for the boundary conditions imposed by the
presence of the $B$ field. The endpoints are then not distinguished by
non-commutativity. The second one is to assume that $B_{ij}$ vanishes but that
the string endpoints are constrained to move on a \hbox{\emph{D}1-brane}
located at $x_0^a$ on which a constant magnetic field $F_{12}=B$ is defined.
The boundary conditions for $X^1$ and $X^2$ then remain unchanged while those
for $X^a$ become $X^a\big|_{\sig=0,\pi}= x^a_0$.  The only difference with the
situation discussed here is that the mode expansion for $X^a$ is no
longer~(\ref{Xa}) but rather
\begin{equation*}
  X^a(\tau,\sig) = x^a_0 + \sum_{n\neq 0} i\>\frac{c_a^a}{n}~\sin n\sig~
     e^{-in\sig}\,.
\end{equation*}
This only introduces some trivial modifications in the analysis of the flat
$a$-directions~\cite{Chu-Ho-1}. From this point of view, the plane-wave metric
extends non-commutativity outside the \hbox{\emph{D}1-brane}.  

\section{The Fock-Hilbert space and the spectrum}

We want to solve the eigenvalue problem 
\begin{equation*}
  H|\psi\rangle = E|\psi\rangle\,,
\end{equation*}
where the hamiltonian is the integral over $\sig$ of the hamiltonian density
${\cal H}$ in equation~(\ref{ham-den}). As discussed in Subsections 2.1 and
2.3, the classical hamiltonian is not positive. This translates, upon
quantization, into an unbounded hamiltonian operator from below. It will
become explicit below that it is precisely the modes with imaginary $\la$ that
make the hamiltonian unbounded, as otherwise was to be expected. Hence not all
the states to be constructed in this Section are within reach for an observer
but only those with positive eigenenergies.

It is convenient to split $H$ as the
sum
\begin{equation*}
  H = H_{\rm flat} + H_{pp}
\end{equation*}
of a contribution 
\begin{equation*}
   H_{\rm flat} = \frac{1}{4\pi\a^\prime}\int_0^\pi \!d\sig \, \Big[\, 
           {\big(\pa_\tau X^a\big)}^2
         + {\big(\pa_\sig X^a\big)}^2 \,\Big] 
\end{equation*}
from the flat $a$-directions and a contribution
\begin{equation*}
   H_{pp} = \frac{1}{4\pi\a^\prime} \int_0^\pi \!d\sig \,\Big\{
     {\big( \pa_\tau X^i\big)}^2
     + {\big(\pa_\sig X^i\big)}^2 
       - m^2\ka^2 \big[ {(X^1)}^2 -{(X^2)}^2 \big]\Big\}\,.
\end{equation*}
from the 1,2-directions.  The eigenstates of $H$ are then of the form
$\,|\psi\rangle = |\psi_{\rm flat}\rangle \otimes |\psi_{pp}\rangle\,$ and the
eigenenergies read $\,E=E_{\rm flat} + E_{pp}$, with $\,\{|\psi_{\rm
  flat}\rangle, E_{\rm flat}\}\,$ and $\,\{|\psi_{pp}\rangle,E_{pp}\}\,$ the
solutions to the eigenvalue problems
\begin{align*}
  H_{\rm flat} |\psi_{\rm flat}\rangle & 
          =  E_{\rm flat}|\psi_{\rm flat}\rangle  \\[3pt]
  H_{pp} |\psi_{pp}\rangle & =  E_{pp}|\psi_{pp}\rangle \,.
\end{align*}

\subsection{Eigenvalue problem for $\boldsymbol{H_{\rm flat}}$}

Apart from the number of dimensions, it is the same problem as for the open
string in Minkowski spacetime.  Using the mode expansions for $X^a$, one
obtains for $H_{\rm flat}$ a sum
\begin{equation*}
   H_{\rm flat} = \frac{1}{2\a^\prime}\,  \sum_{n=1}^\infty  
              {\mathbf :}\, c^a_n{}^\dagger c^a_n\ {\mathbf :}
      \,+\, \a^\prime p_a^2 + \, \frac{D-4}{24} 
\end{equation*}
of harmonic oscillator hamiltonians, one for every frequency $n>0$ in every
direction $a$. As usual, $\,:\!AB\!:\,$ denotes normal ordering of $AB$
and the sum $\,\sum\nolimits_{n>0}n\,$ entering the normal ordering constant
has been regulated using $\zeta$-regularization, so that it takes the value
$\zeta(-1)=-1/12$.  The solution for $H_{\rm flat}$ is well known. The Fock
space is formed by states
\begin{equation}
  |\psi_{\rm flat}\rangle = |\psi_{\{k^a_n\}}\rangle 
     = \bigotimes_{a=3}^{D-2} \big| {\{k_n^a\}}_{n=1}^\infty ,p_a\big\rangle 
     \qquad  k_n^a = 0,1,2\ldots \,,
\label{psi-flat}
\end{equation}
with $k_n^a$ the occupancy number of the harmonic oscillator of frequency $n$
in the $a$-direction. The energies of these states are
\begin{equation*}
   E_{\rm flat} = E_{\{k_n^a\}} = \sum_{a=3}^{D-2} \sum_{n=1}^\infty 
      n\;\!k_n^a + \a^\prime p_a^2\, + \, \frac{D-4}{24} \,.
\end{equation*}
We note that the sum over $n$ is actually finite, since for every eigenstates
there is a finite number of non-zero occupancy numbers $k_n^a$. The action
of~$\,c^b_{n}{}\!^\dagger$ and $\,c^b_{n}\,$ on $\,|\{k_r^a\},p_a\rangle\,$ is
$\,\sqrt{2\a^\prime n}$ times the usual one of creation and annihilation
harmonic oscillator operators.

\subsection{Eigenvalue problem for  $\boldsymbol{H_{pp}}$} 

Employing the mode expansions for $X^i$, we obtain after some work that
\begin{equation}
   H_{pp} = H_{\La_\pm} + \dl_{m^2\ka^2,\rm even}~ H_{\{(k,l)\}} 
          + \dl_{m\ka,\rm even}~ H_{\rm e} + \dl_{m\ka,\rm odd}~ H_{\rm o}\,,
\label{hamiltonian-pp}
\end{equation}
where $H_{\La_\pm}$ is given by
\begin{equation}
   H_{\La_\pm} = \frac{1}{2\pi\a^\prime}\!\sum_{\la\in\La_\pm} 
          \la\,f(\la)\, c_\la\,c_{-\la}
\label{hamiltoniana-La-pm}
\end{equation}
and $H_{\{(k,l)\}}\, H_{\rm e}\,$ and $ H_{\rm o}$ take the form
\begin{eqnarray}
   &{\ds H_{\{(k,l)\}} =  \frac{1}{4\a^\prime}\sum_{\la(k,l)}\,\la\> 
   \Big[ f_a(\la)\> a_{\la}\>a_{-\la} 
       +  f_b(\la)\> b_{\la}\>b_{-\la} \Big]  }& 
   \label{hamiltonian-square}\\[3pt]
   &{\ds H_{\rm e,o} 
    =  \frac{1}{4\pi\a^\prime}\, 
     \Big[ \> \frac{\pi}{4}~ \cosh(m\ka\pi) 
         + \frac{B^2}{m\ka}~\sinh(m\ka\pi) - 1 \,\Big]\,b_{\rm e,o}^2\,. 
   }& \label{hamiltonian-even-odd}
\end{eqnarray}
We first study the problem for $H_{\La_\pm}$ and postpone the solution for the
pathological modes $m^2\ka^2={\rm even},\,m\ka={\rm even},\,m\ka={\rm odd}$.

We recall that an infinite number of the modes $\,\la\!\in\!\La_\pm\,$ have
real $\la$ and that a finite number of them have imaginary $\la$. We separate
their contributions $H_{\rm R}$ and $H_{\rm I}$ to $H_{\La_\pm}$ and write
\begin{equation*}
   H_{\La_\pm} = H_{\rm R} + H_{\rm I}\,.
\end{equation*}
The eigenstates and eigenvalues of $H_{\La_\pm}$ are
$\,|\psi_{\La_\pm}\rangle\! = |\psi_{\rm R}\rangle \otimes |\psi_{\rm
  I}\rangle\,$ and $\,E_{\La_\pm}\!=E_{\rm R} + E_{\rm I}$, with
$\,\{|\psi_{\rm R}\rangle,E_{\rm R}\}\,$ and $\,\{|\psi_{\rm I}\rangle,E_{\rm
  I}\}\,$ solutions to the problems
\begin{align*}
   H_{\rm R} |\psi_{\rm R}\rangle & =  E_{\rm R}|\psi_{\rm R}\rangle \\[3pt]
   H_{\rm I} |\psi_{\rm I}\rangle & =  E_{\rm I}|\psi_{\rm I}\rangle \,.
\end{align*}

\subsubsection{Solution for $\boldsymbol{H_{\rm R}}$}

The commutation relations~(\ref{commutators-12}) for the operators $c_\la$,
yield for $H_{\rm R}$
\begin{equation*}
   H_{\rm R} = \frac{1}{\pi\a^\prime} 
     \sum_{{\ssc \la\in\La_\pm}\atop{\ssc {\rm Re}\,\la >0}} \!
     \la\,f(\la)\> {\mathbf :}\, c_\la^\dagger c_\la {\mathbf :}\,
            + \,K_{\rm R} \,,
\end{equation*}
where $K_{\rm R}$ is the normal ordering constant
\begin{equation}
   K_{\rm R} = - \frac{1}{2} 
   \sum_{{\ssc \la\in\La_\pm}\atop{\ssc {\rm Re}\,\la >0}} \!\la\,.
\label{K-real}
\end{equation}
The hamiltonian $H_{\rm R}$ is a sum of harmonic oscillators, one for every
real $\,\la\!>\!0$.  The eigenstates of $H_{\rm R}$ are then harmonic
oscillator states
\begin{equation}
   |\psi_{\rm R}\rangle 
        = \big| \{k_\la\}_{{\rm Re}\,\la>0} \big\rangle \qquad
  k_\la = 0,1,2\ldots \,,
\label{psi-real}
\end{equation}
with $k_\la$ the occupancy number for the harmonic oscillator of frequency
$\la$, while the eigenenergies read 
\begin{equation*}
  E_{\rm R} = E_{\{k_\la,{\rm Re}\la>0\}} 
     = \sum_{{\rm Re}\,\la >0} \!\la\, k_\la  + \,K_{\rm R} \,.
\end{equation*}
The action of $c_\la^\dagger$ and $c_\la$ on the states
$\,|\{k_{\la^\prime}\}\rangle\,$ is $\sqrt{\pi\a^\prime/f(\la)}\,$ times the usual
action of annihilation and creation harmonic oscillator operators.

Since there are infinitely many positive real $\la$ with no accumulation
point, the normal ordering constant $K_{\rm R}$ needs regularization. For
every $m\ka$ we can always take a sufficiently large integer $N$ such that
$\,m^2\ka^2\!\ll \!N^2\,$ and split the sum for $K_{\rm R}$ into two sums: one
over $0\!<\!\la\!<\!N$ and one over $N\!<\!\la$.  Since $(m\ka/N)^2\!\ll\!
1$, the $\la'{\rm s}$ in the second sum are given by
equation~(\ref{eigen-right}), so that $K_{\rm R}$ can be written as
\begin{equation*}
   K_{\rm R} = -\frac{1}{2} \sum_{{\rm Re}\la< N}\!\!\la\, 
      + \frac{1}{2}\sum_{n=1}^{N} \big( \la_n + \tilde{\la}_n\big)\,
      - \frac{1}{2}\sum_{n=1}^\infty\big( \la_n + \tilde{\la}_n\big)\,.
\end{equation*}
The first two terms in this equation are finite, while accordingly
to~(\ref{eigen-right}) the third one contains the divergent sum
$\,\sum\nolimits_{n>0} n$.  Regularizing this in the same way as for the flat
$a$-directions we arrive at
\begin{equation*}
   K_{\rm R} = \frac{1}{12} +  \Delta K(m)\,,
\end{equation*}
where $\Delta K(m)$ collects all $m$-dependent contributions to $K_{\rm R}$.
For example, for $m^2\ka^2\!\ll\! 1$ the integer $N$ can be taken equal to 1
and from Section 2 it is straightforward to see that
\begin{equation*}
  \Delta K(m^2\ka^2\ll 1) = - \frac{m\ka}{\,2\sqrt{1+B^2}\,}\bigg[
  1 - \frac{(m\ka)^2}{12}~\frac{\pi^2 B^2}{1+B^2}
  + \,{\cal O} \big(m^3\ka^3\big) \bigg]\,.
\end{equation*}

\subsubsection{Solution for $\boldsymbol{H_{\rm I}}$}

It is convenient to introduce for every imaginary $\la$ operators
$\hat{q}_\la$ and $\hat{p}_\la$ defined by
\begin{equation}
  c_{\pm\la} = \sqrt{\frac{\pi\a^\prime}{2\,|\la f(\la)|}}~
        \big(\hat{q}_\la \pm \hat{p}_\la\big)  \qquad {\rm Im}\,\la>0\,.
\label{new}
\end{equation}
They are hermitean and satisfy commutation relations
$\,[\hat{q}_\la,\hat{p}_\la] = i\,{\rm sign}\big[\la f(\la)\big]$. In terms of
them, $H_{\rm I}$ takes the form
\begin{equation*}
   H_{\rm I} = \!\sum_{{\ssc\la\in\La_\pm}\atop{\ssc{\rm Im}\,\la >0}} \!
        {\rm sign}\big[\la f(\la)\big]\>
        \big(\, \hat{p}_\la^2 - \hat{q}_\la^2 \,\big)\,.
\end{equation*}
It is clear that the $H_{\rm I}$ is not bounded from below.  Let us forget for
a moment about this and formally solve the eigenvalue problem for $H_{\rm I}$.
The solution is given by $\,|\psi_{\rm I}\rangle\! =
\prod|\varphi_\la\rangle\,$ and $\,E_{\rm I}\!= \sum E_\la\,$, with the
product and the sum extended over all imaginary $\la$ with $\,{\rm
  Im}\la\!>\!0$, and $\{|\psi_\la\rangle, E_\la\}$ being solutions of
\begin{equation}
   \big(\,\hat{p}_\la^2 - \hat{q}_\la^2\, \big) 
   |\psi_\la\rangle = E_\la |\psi_\la\rangle  \qquad {\rm Im}\,\la>0\,.
\label{eigenproblem-im-one}
\end{equation}
To solve~(\ref{eigenproblem-im-one}) we work in a position representation, in
which the wave function for $|\psi_\la\rangle$ is $\psi_\la(q_\la)$ and the
operators $\hat{q}_\la$ and $\hat{p}_\la$ act on it through multiplication and
derivation, i.e. $\,\hat{q}_\la\to q_\la\,$ and $\,\hat{p}_{\la}\to i\frac{\ds
  d}{\ds dq_\la}$. Equation~(\ref{eigenproblem-im-one}) then becomes
\begin{equation*}
   \Big( \frac{d^2}{d q^2_\la} + q^2_\la + E_\la \Big) \,\psi_\la(q_\la) 
        = 0  \qquad {\rm Im}\,\la>0\,.
\end{equation*}
This is the time-independent Schr\"odinger equation for a particle in an
inverted harmonic potential. Such equation does not have bound states and for
every real $E_\la$ admits
\begin{equation*}
   \psi_{\la,1}(q_\la)= e^{-iq^2_\la/2}\, q_\la\, 
       \Phi\big( \tfrac{3}{4}+\tfrac{iE}{4}\,,\tfrac{3}{2}\,;iq^2_\la\big)
\end{equation*} 
and
\begin{equation*}
 \psi_{\la,2}(q_\la)= e^{-iq^2_\la/2}\,\Phi\big(
      \tfrac{1}{4}+\tfrac{iE}{4}\,,\tfrac{1}{2}\,;iq^2_\la\big)
\end{equation*}
as two linearly independent solutions, $\,\Phi(\mu,\nu;z)\,$ being the
degenerate hypergeometric function.  Both $\,\psi_{\la,1}(q_\la)\,$ and
$\,\psi_{\la,2}(q_\la)\,$ are regular at $q_\la=0$, while at
$|q_\la|\to\infty$ are superpositions of the oscillating exponentials
\begin{equation*}
   \frac{1}{\sqrt{|q_\la|}} ~\, {\rm exp}\,\Big[ \pm \frac{i}{4}\, 
       \big(E_\la \ln q_\la^2 + 2 q_\la^2\big)\Big]\,.                
\end{equation*}
The most general solution for $\psi_\la(q_\la)$ is then an arbitrary linear
combination
\begin{equation*}
    \psi_\la(q_\la)= C_1\, \psi_{\la,1}(q_\la) + C_2\, \psi_{\la,2}(q_\la)\,.
\end{equation*}

The state $\,\psi_\la(q_\la)\,e^{iE_\la\tau}\,$ is a scattering state which in
this position representation is asymptotically formed by one incoming and one
outgoing traveling wave. It is worth noting that these waves are not plane and
that the effect of the inverted harmonic potential is felt at
$|q_\la|\to\infty$. The eigenstates of $H_{\rm I}$ are then
\begin{equation}
      |\psi_{\rm I}\rangle 
         = \big|\{E_\la\}_{{\rm Im}\la>0} \big\rangle
          \to \prod_{{\rm Im}\la>0} \psi_\la(q_\la) 
      \qquad E_\la~\text{real and arbitrary,}  
\label{psi-im} 
\end{equation}
and the energies read 
\begin{equation*}
   E_{\rm I}=\sum_{{\rm Im}\la>0}  {\rm sign}\big[\la f(\la)\big]\>
   E_\la\,.
\end{equation*}
The action of $c_{\pm\la}$ on $\psi_\la(q_\la)$ is through~(\ref{new}) and
multiplication and derivation. The states $|\psi_{\rm I}\rangle$ play in the 1
and \hbox{2-directions} the equivalent r\^ole to that of the plane wave
states~$|p_a\rangle$ in the flat $a$-directions. One way to ensure that the
eigenenergies are non-negative is to restrict to scattering states with
$E_\la= {\rm sign}\big[\la f(\la)\big]\,|E_\la|$ for every imaginary $\la$.

Putting everything together, the eigenstates and
eigenvalues of $H_{\La_\pm}$ are
\begin{eqnarray*}
  & {\ds |\psi_{\La_\pm}\rangle =|\psi_{\{k^a_n\}}\rangle \otimes|\{k_\la\}_{{\rm
      Re}\la>0}\}\rangle \otimes |\{E_\la\}_{{\rm Im}\la>0}\rangle } & \\
  & {\ds E_{\La_\pm}\! = \sum_{a=3}^{D-2} \sum_{n=1}^\infty n\;\!k_n^a 
     + \sum_{{\rm Re}\,\la >0} \!\la\, k_\la 
     + \a^\prime p_a^2\, 
     + \sum_{{\rm Im}\la>0} {\rm sign}\big[\la f(\la)\big]\,E_\la 
     + \, \frac{D-2}{24} + \,\Delta K(m)\,.} &
\end{eqnarray*}

\subsubsection{Contributions from  $\boldsymbol{H_{\{(k,l)\}},\,H_{\rm e},\,H_{\rm o}}$}

If $m\ka$ is such that it squares to an even integer, or is itself an even or
odd integer, the hamiltonian also receives the contributions
$H_{\{(k,l)\}},\,H_{\rm e}$ and $H_{\rm o}$
in~(\ref{hamiltonian-square})-(\ref{hamiltonian-even-odd}). In case
$m^2\ka^2\!=\!{\rm even}$, it is trivial to see that
\begin{equation*}
    H_{\{(k,l)\}} = \frac{1}{2\a^\prime} \sum_{\la(k,l)>0} 
     \la \,\big[ \, f_a(\la) \>a^\dagger_{\la}\,a_{\la} 
               + f_a(\la) \>b^\dagger_{\la}\,b_{\la} \, \big]
     +\sum_{\la(k,l)>0} \la\,.
\end{equation*}
This only adds to the total hamiltonian two harmonic oscillators for every
$\,\la(k,l)$, one for the $\,a_{\la}$-mode and one for the $\,b_{\la}$-mode,
and contributes to the normal ordering constant with a finite quantity. The
eigenstates and eigenenergies are trivial to write.  Assume for example
$m^2\ka^2=6$. There is then only one solution for $(k,l)$, namely $k=2,\,l=1$
and $\la=\sqrt{10}$. This adds to oscillators to the total
hamiltonian and $\sqrt{10}$ to the normal ordering constant. 

For $m\ka\!=\!{\rm even}$ and $m\ka\!=\!{\rm odd}$, the hamiltonian adds an
$m$-dependent momentum-like contribution to the energy.

\section{Conclusion and outlook}

In this paper we have canonically quantized the open string on the Penrose
limit of $dS_n\times S^n$ supported by constant antisymmetric $B_2$ and a
constant dilaton.  Canonical quantization has proved perfectly suited for the
task, thus making unnecessary to resort to Dirac quantization and avoiding the
problem of whether the boundary conditions for the string endpoints should be
regarded as first or second class constraints.  The position operators for the
quantized string define non-commutative spaces, the wave fronts, for all
values of the string parameter $\sig$. Noticeably non-commutativity is not
restricted to the string endpoints but extends outside the brane on which the
endpoints may be assumed to move. The Minkowski limit is smooth and reproduces
the results in the literature~\cite{Chu-Ho-1}.

We think that further investigation of strings on plane-wave backgrounds is
worth to understand non-commutativity in relation with gravity.  The
low-energy field-theory limit looks particularly interesting since it may shed
light on an effective theory for non-commutative gravity.  It must be
mentioned in this regard that there is a vast literature~\cite{Szabo} on the
formulation of Seiberg-Witten maps for gravity and effective non-commutative
corrections to general relativity solutions, plane waves among them~\cite{MR}.

From a purely string theory point of view, the strings considered here may be
thought of as ``in'' or ``out'' states to study string scattering on more
complicated spaces, which in turn will have a Penrose limit, and strings near
spacetime singularities~\cite{DeVega}.

\section*{Acknowledgments}

The authors are grateful to MEC and CAM, Spain for partial support through
grants Nos.  FIS2005-02309 and UCM-910770. The work of GHR was supported by an
MEC-FPU fellowship.

\section*{Appendix A. Explicit expression for the string momentum}
\renewcommand{\theequation}{A.\arabic{equation}}

We collect here the contributions to the string momentum components
$\,P_i(\tau,\sig)\,$ in equation~(\ref{solution-P}) of the various existing
modes. They are obtained by using~(\ref{Pi}) for $X^i_{\rm odd},\, X^i_{\rm
  even}$, $X^i_{(k,l)}$ and $\,X^i_\la$.  For the modes $X^i_{\rm o}\,$ and
$\, X^i_{\rm e}\,$, in~(\ref{zero-1-odd})-(\ref{zero-2-odd})
and~(\ref{zero-1-even})-(\ref{zero-2-even}), we have
\begin{align*}
   2\pi\a^\prime\,P_{1,{\rm o}} &= - \,\frac{m\ka}{B}~b_{\rm o} \>\bigg\{
    \sinh\!\Big( \frac{m\ka\pi}{2}\Big) \cos(m\ka\sig) 
    - B^2\sinh\!\Big[ m\ka\Big(\frac{\pi}{2}-\sig\Big)\Big]\bigg\} \\[4.5pt]
   2\pi\a^\prime\,P_{2,{\rm o}} &= -\, m\ka B \>\bigg[\, a_{\rm o} 
     - \frac{m\ka}{B}~b_{\rm o}\tau\sinh\!\Big( \frac{m\ka\pi}{2}\Big)\bigg]
       \sin(m\ka\sig)
\end{align*}
and
\begin{align*}
   2\pi\a^\prime\,P_{1,{\rm e}} & = \frac{m\ka}{B}~b_{\rm e} \>\bigg\{
     \cosh\!\Big( \frac{m\ka\pi}{2}\Big) \cos(m\ka\sig)    
    + B^2 \cosh\!\Big[ m\ka\Big(\frac{\pi}{2}-\sig\Big)\Big]\bigg\} \\[4.5pt]
   2\pi\a^\prime\,P_{2,{\rm e}} & = -\,m\ka B \> \bigg[\, a_{\rm e} 
    + \frac{m\ka}{B}~b_{\rm e}\tau \cosh\Big(\frac{m\ka\,\pi}{2} \Big)\, 
     \bigg] \sin(m\ka\,\sig)\,.
\end{align*}
The contribution of the modes $\,X^i_{(k,l)}\,$
in~(\ref{particular-1})-(\ref{particular-2}) in turn reads
\begin{align*}
   2\pi\a^\prime\,P_{1,(k,l)} & =  \bigg[ \,
      \frac{\a}{B}~a_{\la(k,l)}\,\Big( \cos\b\sig +B^2\cos\a\sig\Big)  
    + b_{\la(k,l)}\,\big(\sin\b\sig - \frac{\a\b}{\la^2}~\sin\a\sig\Big)
      \bigg]\, e^{-i\la\tau} \\[4.5pt]
   2\pi\a^\prime\,P_{2,(k,l)} & = i\la\,\bigg[ 
      a_{\la(k,l)} \,\Big( \sin\a\sig - \frac{\a\b}{\la^2}~\sin\b\sig\Big)
    + \frac{\b}{B\la^2}~b_{\la(k,l)}\, \Big(\cos\a\sig + B^2\cos\b\sig\Big)\,
      \bigg]\, e^{-i\la\tau} \,.
\end{align*}
Finally, the modes $X^i_\la$ in (\ref{non-zero})-(\ref{X2-sol}) yield the
contributions
\begin{align}
    2\pi\a^\prime P_{1,\pm}\,(\tau,\sig) = 
      c_\la~\frac{\a}{B}\, &\, \bigg[\>
         \cos\b\sig + \frac{\sin\b\pi}{\cos\b\pi\mp 1} ~\sin\b\sig  
       \nonumber \\
         & - B^2\, \Big(\frac{\cos\a\pi\pm 1}{\sin\a\pi}~ \sin\a\sig 
             - \cos\a\sig \Big) \bigg] \, e^{-i\la\tau} \nonumber
 \end{align}
 and
 \begin{align}
    2\pi\a^\prime P_{2,\pm}\,(\tau,\sig) = i\,c_\la\,\la 
     & \bigg[\> \frac{\cos\a\pi\pm 1}{\sin\a\pi}~\cos\a\sig 
                      + \sin\a\sig \label{P1} \\
   & - \frac{\a\b}{\la^2}\> \Big( \sin\b\sig 
         -\frac{\sin\b\pi}{\cos\b\pi\mp 1}~\cos\b\sig\Big) \bigg] \, 
        e^{-i\la\tau}\,. \label{P2}
\end{align}

\section*{Appendix B. Derivation of eqs.~(\ref{Th-ends})-(\ref{Th-inner})} 
\renewcommand{\theequation}{B.\arabic{equation}}
\setcounter{equation}{0}

Organizing the modes in the four sets $\,\La^{\rm I},\, \La^{\rm R},\,
\La,\,\tilde{\La}\,$ introduced in Section 3 and expanding~(\ref{Th-general})
in powers of $m^2\ka^2\ll 1$, the function $\Theta(\sig,\sig^\prime)$ becomes
a sum
\begin{equation*}
  i\,\Theta(\sig,\sig^\prime) = \sum_{k=I,R,\La,\tilde{\La}} 
    i\, \Theta(k;\sig,\sig^\prime)
\end{equation*}
of four contributions $\Theta(k;\sig,\sig^\prime)$, each one of which is a
power series in $m\ka$. Up to order four in $m\ka$, these contributions read
\begin{align}
   i\,\Theta({\rm I};\sig,\sig^\prime) & =  
    \, \frac{i\a^\prime B}{\,2\,(1+B^2)\,}~ (\pi - 2\sig) 
    \label{Th-I}\\[3pt]
    & - \,\frac{\,i\a^\prime B\,(m\ka)^2}{\,12\,(1+B^2)^2\,}~ 
     \big[ \sig\,(\sig-\pi)\,(2+B^2) 
        - 3B^2\sig^\prime\,(\sig^\prime-\pi) -\pi^2 \big] \,,
    + \,{\cal O}\big(m^4\ka^4)  \nonumber
\end{align}
\begin{align}
    i\,\Theta({\rm R};\sig,\sig^\prime) & = 
      \, \frac{i\a^\prime B}{\,2\,(1+B^2)\,}~ (\pi -2\sig^\prime) 
      \label{Th-R}\\[3pt]
    & + \, \frac{\,i\a^\prime  B (m\ka)^2}{\,12\,(1+B^2)^2\,}~ \big[ 
         \sig^\prime\,(\sig^\prime-\pi)\,(2+B^2) 
        - 3B^2\sig\,(\sig-\pi) - \pi^2 \big] 
      + \,{\cal O}\big(m^4\ka^4)  \,, \nonumber
\end{align}
\begin{align}
   i\,\Theta(\La;\sig,\sig^\prime) & = -\,\frac{2i\a^\prime B}{\,1+B^2\,} 
     \sum_{n=1}^\infty ~\frac{\cos n\sig \sin n\sig^\prime }{n} 
      \label{Th-La} \\[3pt]
  & + \, \frac{\,i\a^\prime B (m\ka)^2\,}{(1+B^2)^2}\>\sum_{n=1}^\infty
      \bigg[\, B^2 (2\sig-\pi) \> \frac{\sin n\sig \sin n\sig^\prime}{n^2}  
       + (2\sig^\prime -\pi)\> \frac{\cos n\sig \cos n\sig^\prime}{n^2} 
       \nonumber \\[3pt]
  & 
  \hphantom{+ \, \frac{\,i\a^\prime B\,(m\ka)^2\,}{(1+B^2)^2}
      \>\sum_{n=1}^\infty\bigg[ } 
      - 2\,(1-B^2)~ \frac{\cos n\sig\,\sin n\sig^\prime}{n^3}\>
       \bigg]  + \,{\cal O}\big(m^4\ka^4) \nonumber
\end{align}
and 
\begin{align}
   i\,\Theta(\tilde{\La};\sig,\sig^\prime) = & - 
      \frac{2i\a^\prime B}{\,1+B^2\,}~ \sum_{n=1}^\infty~
     \frac{\sin n\sig \cos n\sig^\prime}{n} \label{Th-tilde} \\[3pt]
   & - \frac{\,3i\a^\prime B\,(m\ka)^2}{(1+B^2)^2}\sum_{n=1}^\infty~ 
       \frac{\sin n\sig \cos n\sig^\prime}{n^3} ~\nonumber
       + \,{\cal O}\big(m^4\ka^4)\,.
\end{align}
Summing all the contributions of order zero in $m\ka$ in these equations, we
have
\begin{equation*}
   i\Theta_0(\sig,\sig^\prime) = \frac{i\a^\prime B}{1+B^2}~
     \big[ \pi -\sig_+ - F_1(\sig_+) \big]\, 
\end{equation*}
where $F_1(\sig_+)$ is the Fourier series~(\ref{F2}). This trivially
leads to the order zero contributions in eqs.~(\ref{Th-ends})
and~(\ref{Th-inner}). To sum the order two contributions, we first note that
\begin{align}
   \big[i\Theta(\La)  +  i\Theta(\tilde{\La})\big]_2 = & - 
     \frac{i\a^\prime B}{4\,(1+B^2)^2}~ \Big\{
       \big[\, B^2(\sig_+ + \sig_- -\pi) 
           + (\sig_+ -\sig_- -\pi)\, \big]\,F_2(\sig_-) \nonumber\\[3pt]
  & - \big[\, B^2(\sig_+ + \sig_- -\pi) 
           - (\sig_+ -\sig_- -\pi)\, \big]\,F_2(\sig_+) \nonumber \\[3pt]
  & + (2B^2+1)\,F_3(\sig_+) + ( 5-2B^2)\,F_3(\sig_-) \Big\} \,,
  \label{Th-aux}
\end{align}
where the Fourier series $F_2(\sig_\pm)$ are as
in~(\ref{F3})-(\ref{F4}) and $F_3(\sig_\pm)$ read
\begin{align*}
   F_3(\sig_-):= 2 \sum_{n=1}^\infty \frac{\sin n\sig_-}{n^3} 
     & = \frac{\sig_-^3}{6} 
       - \frac{\pi}{2}\>\sig_-\,|\sig_-| + \frac{\pi^2}{3}\>\sig_- \\[6pt]
   F_3(\sig_+) := 2 \sum_{n=1}^\infty \frac{\sin n\sig_+}{n^3} 
     & = \frac{\sig_+^3}{6} 
       - \frac{\pi}{2}\>\sig_+^2 + \frac{\pi^2}{3}\>\sig_+ \,.
\end{align*}
Eqs.~(\ref{Th-I}), (\ref{Th-R}) and~(\ref{Th-aux}) then lead to the second
order contributions in Section~5.

\end{document}